\documentclass[usenatbib]{mn2e}
\usepackage{epsfig}
\usepackage{graphicx}
\usepackage{ifthen}
\usepackage{amssymb}
\usepackage{subfigure}
\def\kms{\,\rm km\,{s}^{-1}}

\def\beq{\begin{equation}}
\def\eeq{\end{equation}}

\def\omega0{\Omega_{\rm m,0}}
\def\kpc{\,\rm kpc}
\def\mpc{\,\rm Mpc}
\def\LCDM{\Lambda{\rm CDM}}
\def\zs{z_{\rm s}}
\def\zl{z_{\rm l}}
\def\betal{\beta_{l}}
\def\betah{\beta_{h}}

\def\Deltam{{\Delta m}}
\def\Nimage{{N_{\rm img}}}

\newcommand\apj{{ApJ}}% % Astrophysical Journal 
\newcommand\aap{{A\&A}}% % Astronomy and Astrophysics 
\newcommand\mnras{{MNRAS}}% % Monthly Notices of the RAS 
\newcommand\nat{{Nature}}% % Nature \newcommand\aas{{AAS}}% 
\date{
Accepted 2007 April 3.
Received 2007 April 3;
in original form 2007 January 28}

\pubyear{2007}
\volume{378}
\pagerange{469-481}
\title{Properties of Wide-separation Lensed Quasars by Clusters
  of Galaxies in the SDSS}
\author[Li et al.]{G.L. Li$^{1,2}$\thanks{E-mail: {\tt lgl@shao.ac.cn}},
 S. Mao$^{3}$,
  Y.P. Jing$^{1,2}$, W.P. Lin$^{1,2}$, M. Oguri$^{4}$
\\
$^{1}$ Shanghai Astronomical Observatory; the Partner
Group of MPA, Nandan Road 80, Shanghai 200030, China
\\
$^{2}$ Joint
Institute for Galaxy and Cosmology (JOINGC) of SHAO and USTC
\\
$^{3}$University of Manchester, Jodrell Bank Observatory,
 Macclesfield, Cheshire SK11 9DL, U.K.
\\
$^{4}$Kavli Institute for Particle Astrophysics and
Cosmology, Stanford University, 2575 Sand Hill Road, Menlo Park, 
CA 94025}
\begin{document}
\maketitle
\begin{abstract}
We use high-resolution $N$-body numerical simulations to study the number of
predicted large-separation multiply-imaged systems produced by clusters
of galaxies in the SDSS photometric and spectroscopic quasar samples.
We incorporate the condensation of baryons at the centre of clusters by
(artificially) adding a brightest central galaxy (BCG)
as a truncated isothermal sphere.
We make predictions in two flat cosmological models:
a $\LCDM$ model with a matter density $\omega0=0.3$, and $\sigma_8=0.9$ ($\LCDM$0), and a
model favoured by the WMAP three-year data with $\omega0=0.238$, and $\sigma_8=0.74$ (WMAP3).
We found that the predicted multiply-imaged quasars with separation
$>10\arcsec$ is about 6.2 and 2.6 for the SDSS photometric (with an effective area 8000 deg$^{2}$) and
spectroscopic (with an effective area 5000 deg$^{2}$) quasar samples respectively in the $\LCDM$0 model;
the predicted numbers of large-separation lensed quasars agree well with the observations.
These numbers are reduced by a factor of 7 or more in the WMAP3 model, and 
are consistent with data at $\la 8\%$ level.
The predicted cluster lens redshift peaks around redshift 0.5, and 90\% are
between 0.2 and 1. The ratio of systems with at least four image systems
($\Nimage\ge 4$) and those with $\Nimage \ge 2$ is about 1/3.5 for both
the $\LCDM$0 and WMAP3 models, and for both the photometric and
spectroscopic quasar samples. We find that the BCG creates a central
circular region, comparable to the Einstein ring of the BCG, where the central
image disappears in the usual three-image and five-image configurations.
If we include four image systems as an extreme case of five-image
systems (with an infinitely demagnified central image), we find that
68\% of the central images are fainter by a factor of 100 than the
brightest image, and about 80\% are within $1.5\arcsec$ of the BCG.
\end{abstract}

\section{INTRODUCTION}

The number of multiply-imaged quasars lensed by galaxies has now reached roughly
one hundred\footnote{http://cfa-www.harvard.edu/castles/}.
The typical separation of these systems ranges from $0.3-6\arcsec$.
They provide an valuable sample to constrain the 
cosmological constant, the Hubble constant, and the 
mass profiles of lensing galaxies at intermediate redshift (for an
extensive review, see 
Kochanek, Schneider \& Wambsganss 2006 and references therein).

The search for multiply-imaged quasars by clusters of galaxies has been less successful. The
initial search for wide-separation radio sources yielded no successful
candidates between $6-15\arcsec$ (\citealt{Phi01a}), and from $15-60\arcsec$
(\citealt{Phi01b}). \citet{Ofek01} also failed to find any wide-separation quasars between $5-30\arcsec$ in
the FIRST 20-cm radio survey. The lack of cluster lensed radio sources
is likely due to the small number of radio sources surveyed ($\sim$ 20,000).
The breakthrough came from the Sloan Digital Sky Survey (SDSS) where a
large number of optical quasars became available. So far, there are approximately
46,420 spectroscopically confirmed broad-line quasars in SDSS Date Release 3. 
Two spectacular cluster lenses were discovered in the SDSS (not limited to DR3, see \S\ref{sec:number}). The first case was SDSS J1004+4112, a
quadruply-imaged system with an separation of 14.6 arcsec
(\citealt{Inada03, Oguri04}). The cluster is at redshift
of 0.68, and the lensed quasar is at redshift 1.734. 
A faint fifth image was later discovered (\citealt{Inada05}), as were 
many other multiply-imaged background galaxies (\citealt{Sharon05}). 
The second case, SDSS J1029+2623, was a double-image system with a separation of 22.5 arcsec.
The lensing cluster is likely at redshift 0.55 (\citealt{Inada06}), and the background quasar at
redshift $\zs=2.197$.

Cluster lenses, just like galaxy-scale lenses, provide 
valuable constraints on the lens mass profiles.
They are also important because the number of such lenses depend quite 
sensitively on the cosmology, in particular the matter power-spectrum
normalisation, $\sigma_8$, and the matter density, $\omega0$. This
sensitivity has already been explored using giant arc statistics where the lensed
sources are background galaxies, rather than quasars
(e.g., \citealt{Mene03a, Mene03b, Wambsganss04, Torri04, Dalal04,
Horesh05, Li05, OLS03}). The analysis by \citet{Li06b} prefers a high $\sigma_8$ universe.
However, the study and other similar ones suffer from one deficiency, namely the properties of
the lensed background galaxy population, in particular
their redshift distribution, is not well known, so the conclusion is somewhat uncertain. 

In comparison, the quasar population has a much better-known 
redshift distribution, and so does not suffer as much from the same deficiency, hence
a careful analysis of the statistics of multiply-imaged quasars is
complementary to those of giant arcs. Of course, the number of known
multiply-imaged quasars so far is very small (two), compared to hundreds
of known giant arcs (e.g., \citealt{Luppino99, Gladders03, ZG03, Sand05, Hen06}).
Nevertheless, even this small sample provides interesting constraints.
For example, the analysis of \citet{Oguri04} of the first cluster lens, J1004+4112, 
concluded that the $\sigma_8$ value must be high,
$\sigma_8=1^{+0.4}_{-0.2}$ (95\% confidence) even for an inner density profile,
$\rho \propto r^{-1.5}$, considerably steeper than the NFW profile
($\rho \propto r^{-1}$ for small $r$, Navarro, Frenk \& White 1997). Many studies of clusters of galaxies
prefer the NFW profile (e.g., \citealt{vdM00, Com06, VF06}). For such a
shallower profile, the required $\sigma_8$ would be even higher.

Another advantage of lensed quasars concerns the lens selection.
For giant arc surveys, observers first select clusters from 
optical or X-ray data and then take deep followup images, hence the discovery
of giant arcs depends on how clusters are selected in the first place. On the other hand, lensed
 quasars are usually identified by examining quasars to see whether they
 are multiply imaged. In other words, current arc surveys only observe
biased directions whereas quasar lenses survey random line-of-sight. In addition,
quasars are point sources, so one does not have complications arising from
seeings, background source size and ellipticity distributions etc, thus it
is much simpler to model the quasar source population.

There have been many previous works to predict
the abundance of large-separation quasar lensed by massive dark halos using simple spherical models 
(\citealt{Maoz97,SRM01,KM01,WTS01,LO02,Rusin02,LO03,Oguri03,HM04,
  LM04,KKM04,Oguri04,Chen04}). 
 In particular, \citet{LM04}, using spherical clusters,
found that wide-separation lensed quasars can be a powerful tool 
to measure $\sigma_8$ and $\Omega_{m,0}$.
Using triaxial halo models, \citet{OK04}
computed the abundance of quasars multiply imaged by clusters analytically and found
that triaxiality significantly increases the expected number of
lenses.
\citet{Hen07b} presented the first predictions of
multiply-imaged quasars using N-body simulated clusters in the a flat cosmology 
with $\omega0=0.3$, $\Omega_{\Lambda,0}=0.7$ with $\sigma_8=0.95$. 
In this paper, we compare the predictions for the cosmology favoured by
the three-year WMAP result with the usual concordance cosmology.
We also make more detailed predictions
for cluster lenses as a function of image multiplicity, and study
the brightness of the central images and their offset from cluster centres.

The paper is structured as follows. In \S2, we discuss the simulations
we use and our lensing methodology. In \S3, we discuss the background
source population, namely the quasar luminosity function. In \S4, we
present the main results of the paper, and we finish in \S5 with a summary
and a discussion of the implications of our results.

\section{NUMERICAL SIMULATIONS AND LENSING METHODOLOGY}

In this paper, we utilise two sets of simulations, one for the `standard'
$\LCDM$ model with $\sigma_8=0.9$, and the other for the cosmological
model given by the recent three-year WMAP data with
$\sigma_8=0.74$ (\citealt{Spergel06}). For brevity, these two models will be referred to as
$\LCDM$0, and WMAP3, respectively, and the corresponding cosmological
parameters are:
\begin{enumerate}
\item $\LCDM$0:~ $\Omega_{\rm m,0}=0.3,\Omega_{\Lambda,0}=0.7,h=0.7,
     \sigma_8=0.9, n=1$;
\item WMAP3: $\Omega_{\rm m,0}=0.238,\Omega_{\Lambda,0}=0.762,
     h=0.73,\sigma_8=0.74, n=0.95$,
\end{enumerate}
where $h$ is the Hubble constant in units of $100\kms\mpc^{-1}$, and
$n$ is the spectral index of the power spectrum.
The assumed initial transfer function in each model was generated
  with {\small CMBFAST} (\citealt{Seljak96}). Notice that both
  $\sigma_8$ and $\Omega_{\rm m,0}$ differ in these two models, and both
  will
impact on the number of multiply-imaged quasars.

\begin{table}
\caption{Parameters for the two cosmological numerical simulations. The columns are 
cosmology, box size, number and mass of dark matter particles, and the softening length.}
\begin{tabular}{lcccc} 
\hline
model & box size & $N_{\rm DM}$ & $m_{\rm DM}$  &  softening \\
      & $(h^{-1}\mpc)$  &       &   $(h^{-1} M_\odot)$          &  ($h^{-1}$\kpc) \\\hline
$\LCDM$0 & 300 &  $512^3$  &  $1.67\times 10^{10}$   &  30  \\
WMAP3 & 300 &  $512^3$     & $1.32\times 10^{10}$  &  30  \\
\hline
\end{tabular}
\label{table:simu}
\end{table}

We use a vectorised-parallel ${\rm P^3M}$ code (\citealt{JS02})
and a {\small PM-TREE} code -- GADGET2 \citep{Springel01, Springel05} to
simulate the structure formation in these models; the details of
the simulations are given in Table 1. Both are $N$-body simulations
which evolve $N_{\rm DM}=512^3$ dark matter particles in a large cubic
box with sidelength equal to $300 h^{-1} \mpc$. The $\LCDM$0
simulation was performed using the ${\rm P^3M}$ code  of Jing \& Suto
(2002) and has been used in \citet{Li05} to study the properties of
giant arcs. We refer the readers to that paper for more details. 
The WMAP3 simulation was carried out with GADGET2, and has been
used by \citet{Li06b} to compare the statistics of giant arcs
in the $\LCDM$0 and WMAP3 models. Similarly, in this work we will use
them to compare the statistics of multiply-imaged quasars in these two 
cosmological models. We return to the issue of cosmic variance in
\S\ref{sec:cosmic_variance}.

Dark matter halos are identified with the friends-of-friends method
using a linking length equal to 0.2 times the mean particle
separation. The halo mass $M$ is defined as the virial mass enclosed
within the virial radius according to the spherical collapse model
(\citealt{KS96}; \citealt{BN98}; \citealt{JS02}).

For any given cluster, we calculate the smoothed surface density maps
using the method of \citet{Li06a}. Specifically, for any line of
sight, we obtain the surface density on a $1024 \times 1024$ grid
covering a square of (comoving) side length of $4h^{-1}\mpc$ centreed
on each cluster. The projection depth is also chosen to be $4h^{-1}\mpc$.
Notice that the size of the region is large enough to include all particles
within the virial radius (in fact, several virial radii for small
clusters). Particles outside this cube
and large-scale structures do not contribute significantly to the
lensing cross-section  (e.g.  \citealt{Li05}; \citealt{Hen07a}), so we
will ignore their contributions here.

Our projection and smoothing method uses a smoothed particle
hydrodynamics (SPH) kernel to distribute the particle mass on a 
three-dimension grid and then the surface density is obtained by integrating
along the line of sight (see \citealt{Li06a} for more detail). In this work, the
number of neighbours used in the SPH smoothing kernel is fixed to be
32. We calculate the surface density along three orthogonal directions for each cluster.
Once we obtain the surface density of a dark matter halo, its lensing
potential, $\phi_{\rm DM}$, can then be easily calculated
using the Fast Fourier Transform (FFT).

Many clusters of galaxies have a very luminous central galaxy.
The effect of such brightest cluster galaxies (BCGs) has been 
quantified by \citet{Hen07b}. For multiply-imaged systems with
separations $\lesssim 20\arcsec$, BCGs can enhance the cross-section by $\sim$
50\%. Following \citet{Hen07b}, we also artficially add a BCG at
the centre of each cluster.  This is obviously a simplification as BCGs
are assembled as a function of time; we briefly return to this issue
in the discussion. The BCG is modelled as a truncated
isothermal sphere. The mass of BCG is taken to be a fixed 
fraction of the host halo, $M_{\rm BCG}=3\times 10^{-3} M_{\rm FOF}$ and its velocity
dispersion is given by
\begin{equation}
\frac{\sigma}{300 \kms}=\left(\frac{M_{\rm FOF}}{10^{15} \rm M_{\odot}}\right)^{2/15}.
\label{eq:sigma}
\end{equation}
The truncation radius is given by $r_{\rm max} = G M_{\rm BCG}/(2\sigma^2)$.
See \citet{Hen07b} for more details.

In the following we will only present results including such a BCG,
which approximately accounts for the effects of baryonic cooling
at the centre of clusters. The effect of
baryons in non-central galaxies is likely to be smaller.
As the BCG is modelled as a truncated isothermal sphere, its lensing
potential, $\phi_{\rm BCG}$, can be calculated analytically.
The total lensing potential is a sum
of the contributions from the (pure) dark matter halo and
the BCG, $\phi=\phi_{\rm DM}+\phi_{\rm BCG}$. The deflection angle ($\vec\alpha$)
and magnification ($\mu$) can be obtained by taking first-order and second-order derivatives with
respect to $x$ and $y$, $\alpha_x=\phi_x$, $\alpha_y=\phi_y$, and
$\mu=[(1-\phi_{xx})(1-\phi_{yy})-\phi_{xy}^2]^{-1}$.
We calculate the BCG contribution to the deflection angle and potential
analytically, and the dark matter halo contribution using FFT as discussed above. 

Using these derivatives, we can create the mapping from the image plane
to the source plane and obtain the critical curves and caustics.
Critical curves are a set of image positions with $\mu=\infty$, and the
caustics are the corresponding source positions.
Caustics play a crucial role in gravitational lensing as they
divide the source plane into regions of different
image multiplicities. Whenever a source crosses a caustic, the image
number either increases or decreases by two. For a quasar to be
multiply-imaged, it must lie inside one of the caustics.

To calculate the cross-section of a given separation of a multiply
imaged quasar, we first find the smallest rectangle which encloses
all the caustics, and then a regular grid is created dividing this
rectangle into pixels of $0.3\arcsec$ by $0.3\arcsec$. 
For each source position located on the grid, we find the corresponding
image positions and magnifications ($\mu$) using the Newton-Raphson
method. For a given multiple-image system, the number of
images one can observe, $\Nimage$, depends
 on the magnification of each image, the source intrinsic magnitude and the
survey magnitude limit. For such systems, we sort the images according to the
image magnifications and record the number of images $\Nimage$ and 
 the maximum separation $\Delta \theta$ between these $\Nimage$ images.
We then examine whether the multiple images satisfy the observational
selection criteria (see \S\ref{sec:k-corr}).
For given lens and source redshifts, $\zl$ and $\zs$, collecting
systems that pass the selection allows us to
construct a cross-section $\sigma(\ge \Nimage, >\theta| \zl, \zs)$ for lensing
systems with at least $\Nimage$ images and separations larger than $\theta$. 

The SDSS photometric and spectroscopic quasar samples are magnitude
limited, multiply-imaged quasars will be over-represented in such
samples due to the lensing magnification (\citealt{Turner80}).
To account for this magnification (or amplification) bias, we further divide the
cross-section into 50 bins of $\Delta m \equiv -2.5 \log \mu$
from $-5$ mag to 5 mag with a bin size of 0.2 mag, and obtain a
differential cross-section $d\sigma/d\Delta m(\ge\Nimage, >\theta|
\Deltam, \zl, \zs)$. 

As in \citet{Li05}, we can obtain the total cross-section per unit
comoving volume by summing over the contributions of all the clusters:
\beq
\overline\sigma(\ge \Nimage, >\theta| \Delta m, \zl, \zs) = 
{\sum \sigma_i(\ge \Nimage, >\theta| \Delta m, \zl,\zs) \over V},
\eeq
where $\sigma_i(\ge \Nimage, >\theta| \Delta m, \zl, \zs)$ is the average cross-section of the 
three projections of the $i$-th  cluster at redshift $\zl$, $\zs$ is the source redshift, and $V$ is
the comoving volume of the simulation box.
Then, accounting for the magnification bias (for more detailed
descriptions, see Sect. 3.1), the predicted number
of large-separation lens can be derived as:
\begin{eqnarray} \label{eq:N}
\nonumber
N(\ge\Nimage, >\theta) =&  \int_0^{\infty} d\zs \int_{0}^{\zs} d\zl \int_{-\infty}^{+\infty}
d\Delta m \,\, \\
\nonumber                       &{d{\overline\sigma} \over d\Deltam} (\ge\Nimage, >\theta| \Deltam, \zl,
\zs) \\ 
               \times & n(<i_0-\Deltam, \zs)\, (1+\zl)^3\, \, {dV_{\rm p}(\zl)\over d\zl} 
\end{eqnarray}
where $n(<i, \zs)\times d\zs$ is the number of quasars brighter than the SDSS
$i$-band magnitude $i$ in the redshift interval
from $\zs$ to $\zs+d\zs$, $i_0$ is the flux limit, 
and $dV_{\rm p}(\zl)$ is
the proper volume of a spherical shell with redshift from $\zl$ to
$\zl+d\zl$. We used 22 simulation output with redshift from 0.1 to 2,
and selected the 600 (for $\LCDM$0) or 400 (for WMAP3) most massive clusters in each output to
calculate the total lensing cross-sections. The lowest masses of clusters in these two
cosmologies are $7.8\times10^{13}h^{-1}M_{\odot}$ and 
$5\times10^{13}h^{-1}M_{\odot}$ respectively at redshift 0.5.
The integration step size of $\zl$ is the same as the redshift interval of
simulation output ($\delta \zl \approx 0.1$) and the source redshift is at 0.7,
1.0, 1.5, 2.0, 2.5, 3.0, 3.5, 4.0, 4.5, 5.0 and 5.5. 

\section{THE QUASAR LUMINOSITY FUNCTION }
\label{sec:lf}
To make accurate predictions for the frequency of multiply-imaged quasars, we must specify
the quasar luminosity function as a function of redshift and
luminosity. Following \citet{Boyle00}, we will use a double power-law to parameterise the quasar
luminosity function
\beq
\Phi(M, z)=\frac{\Phi^{\ast}(z)}{10^{0.4(\beta_{l}+1)[M-M^{\ast}(z)]}
 + 10^{0.4(\beta_{h}+1)[M-M^{\ast}(z)]}},
\label{eq:LF}
\eeq
where $M$ is the absolute magnitude in a filter, and $\Phi^\ast$ and
$M^\ast$ are the characteristic density and magnitudes respectively in a
certain band.

For the low redshift ($z<2.1$) quasars, we use the parameters from the 2dF-SDSS LRG and
QSO (2SLAQ) Survey (\citealt{Richards05}). The luminosity function is fitted
in the SDSS $g$-band with $\betal=-1.45$, $\betah=-3.31$ and 
$\Phi^{\ast}=1.83\times 10^{-6}\rm \,Mpc^{-3} \,mag^{-1}$. The characteristic
luminosity in the $g$-band, $M^\ast_{g}$, is described by a second-order polynomial of redshift:
\beq
M^{\ast}_{g}(z)=M^{\ast}_{g}(0)-2.5(k_1z+k_2z^2),
\eeq
where $M^{\ast}_{g}(0)=-21.61$, $k_1=1.39$ and $k_2=-0.29$.

For high redshift, $z>3.5$, quasars, we take the parameters from \citet{Fontanot06},
which were obtained from a joint analysis of the GOODS and SDSS
data. The best fit to the luminosity function as a function of 
the monochromatic 1450\,\AA\, magnitude, $M_{145}$, gives
$\betal=-1.71$, $\betah=-3.31$; the data also prefers a pure 
density evolution with an exponential form:
\beq
\Phi^{\ast}(z)=\Phi^{\ast}_{(z=2)}e^{k_z[(1+z)-3]},
\eeq
where $k_z=-1.37$ and $\Phi^{\ast}_{(z=2)}=1.67\times 10^{-6}\rm\, Mpc^{-3} \,mag^{-1}$. 
The characteristic monochromatic luminosity, $M^\ast_{145}$, is fixed at $\rm M^{\ast}_{145}=-26.43$.

For a given flux limit $i_0$, we can easily obtain the total number 
density of quasars $\rm n(<i_0,z)$ in the redshift space for $z<2.1$ and 
$z>3.5$. However, the quasar luminosity function is not well constrained from redshift
2.1 to redshift 3.5. For a given magnitude limit $i_0$, we estimate the
number density of quasar with redshift $2.1<z<3.5$ by linear
interpolation as a function of redshift of the quasar number densities
with the same absolute magnitude.
For a given redshift $z$, we first calculate the corresponding absolute
magnitude, $M^0_{i}(z)$ for each apparent magnitude $i$. We then convert this, using
k-corrections (see \S\ref{sec:k-corr}), into corresponding
$g$-band and the 1450\,\AA\, monochromatic absolute luminosities at
$M^0_g$ and $M^0_{145}$ at redshift $z$. Since we know $\Phi(M^0_{g},z=2.1)$, 
$\Phi(M^0_{145},z=3.5)$, we obtain the $\Phi(M_i, z)$ by
linear interpolation:
\beq
\Phi(M_i, z)=\frac{\Phi(M^0_g, 2.1) \times(3.5-z)+ \Phi(M^0_{145}, 3.5)\times(z-2.1)}{3.5-2.1}.
\label{eq:zmiddle}
\eeq
We can then simply integrate the above equation to derive the number density
of quasars per square degree for any limiting magnitude $i$.

We also consider the incompleteness of quasars at redshift from 2.3 to 3.6 as in
\citet{Hen07b}. The SDSS spectroscopic and photometric surveys depend on colour 
selection and the colours of quasars and stars are not easy to disentangle in this redshift range. 
So we assume that only 60\% of quasars in this redshift interval are
found by the SDSS.

\begin{figure}
{
 \centering
% \leavevmode 
\columnwidth=.5\columnwidth
\includegraphics[width={\columnwidth}]{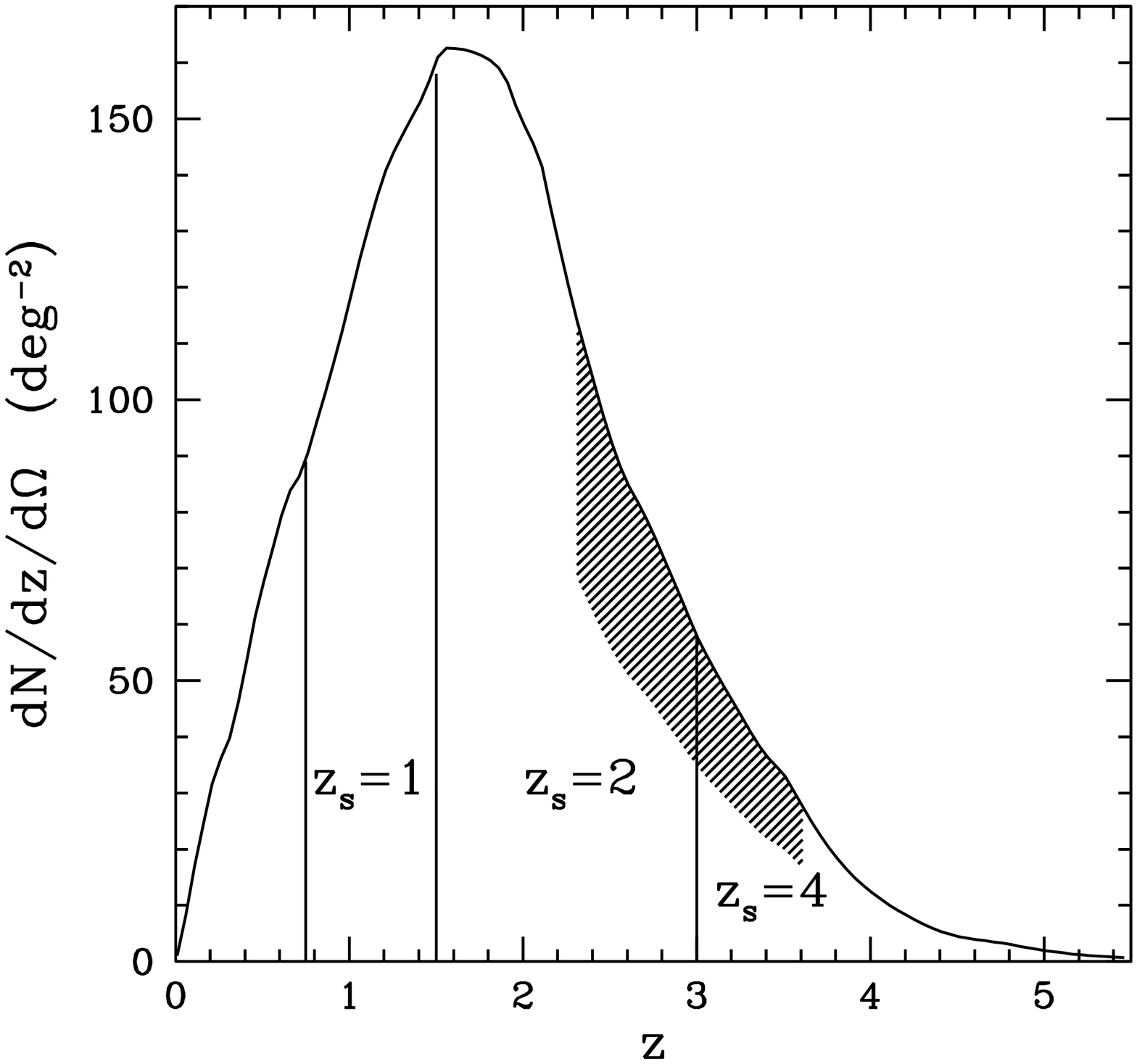}%
\includegraphics[width={\columnwidth}]{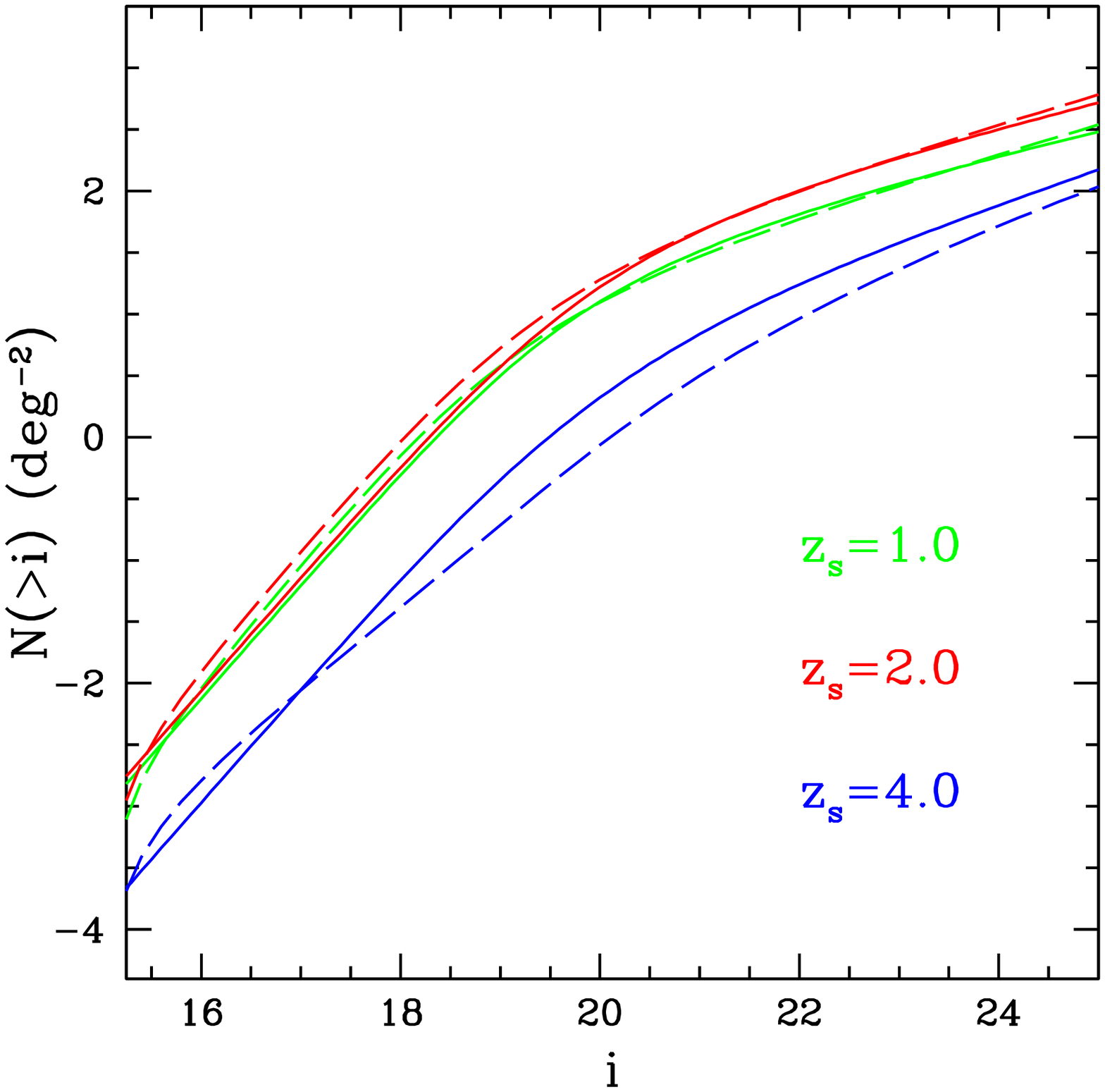}%
}
\caption{The left panel shows the redshift distribution of quasars for a
  flux limit of $i < 22.7$. Three redshift intervals ($\zs\sim 1, 2$ and
  4) are indicated by the vertical lines. The shaded region corresponds to the 60\% completeness we assumed.
 The right panel shows the surface density
  of quasars as a function of the SDSS $i$-band
  magnitude for the three source redshift intervals. The solid and dashed lines
  show the shapes used by us and by \citet{Hen07b}
  respectively. They are in good agreement for the two low redshift intervals, 
but for high reshift interval, our amplitude
  is higher by roughly a factor of 2 than theirs between $i=18$ and 22.
} 
\label{fig:lf}
\end{figure}

\subsection{The cross filter k-correction, quasar samples and lens selection \label{sec:k-corr}}

The SDSS spectroscopic and photometric survey have flux limits in 
the $i$-band but the parameters in the above luminosity functions (in eqs. \ref{eq:LF}-\ref{eq:zmiddle})
are given in different bands, $M_g$ and $\rm M_{145}$. To apply these quasar luminosity
functions to the $i$-band, we have to do cross filter k-corrections (\citealt{Hogg02}).
We use the results of \citet{Richards06} who provided the k-correction 
from the SDSS i-band to the SDSS $g$-band and the monochromatic 1450\,\AA\,
absolute magnitude assuming a power-law quasar spectral index,  $\alpha_{\nu}= -0.5$.

As in \citet{Hen07b}, we use the same photometric and
spectroscopic quasar samples. For the SDSS spectroscopic quasar sample, the covered area is about 5000
$\rm deg^{2}$ and the magnitude limit is $i < 19.1$ for low redshift 
($z < 2.5$) quasars and $i<20.2$ for high redshift ($z > 2.5$) quasars (\citealt{Schneider05}). The faint
SDSS photometric quasar survey covers a larger area, $\sim 8000\, \rm {deg}^2$, 
with a flux limit of $i < 21.0$ for $z < 2.5$ and $i < 20.5$ for $z > 2.5$ 
(\citealt{Richards04}). To select lenses in the photometric and
spectroscopic quasar samples, we require the brightest image to be brighter
than the magnitude limit given above, and other images to be brighter
than $i=21.0$, after accounting for their magnifications.
The fainter limit for other images is because of fiber
collisions in the SDSS, usually only one quasar image will be spectroscopically
targeted, while other quasar candidates have to be
confirmed with other, typically larger telescopes that can reach much
fainter magnitudes. This is what actually occurred for the discovery of
the first two lenses, and so we will adopt this selection criterion,
similar to \citet{Hen07b}. Notice that some images may still be
missed if they are below the flux limit, even after the magnification, so
the observed number of images may be smaller than the
intrinsic number.

Notice that the photometric sample goes almost two
magnitudes deeper than the spectroscopic sample for quasars below
redshift 2.5 (magnitude limit 21.0 vs. 19.1), but only slightly deeper 
for high redshift quasars (20.2 vs. 20.5). As a result, the photometric
sample has many more low-redshift quasars compared with the
spectroscopic sample. This will be reflected in the source and lens
redshift distributions for multiply-imaged quasars
(see Figs. \ref{fig:lens} and \ref{fig:source}).

Fig. \ref{fig:lf} shows the redshift distribution of quasars
 for $i<22.7$. It shows the well known peak around redshift 2, and the sharp decline
beyond redshift 4. The shaded region corresponds to the 60\% completeness we assumed.
This result is in good agreement with 
the left panel of Fig. 1 in \citet{Hen07b}.
% This number
%count distribution as a function of the flux limit $i$ is used in both the
%$\LCDM$0 and WMAP3 cosmologies, as this must match the observed 
%distribution. 
 Notice that the geometries in the $\LCDM$0 and WMAP3 cosmologies 
are different, due to the small difference in the matter density. However,
even for a source at redshift 5, the angular diameter distance differs
only by $<8\%$. Furthermore, we reproduce the quasars distribution
as a function of redshift and magnitude (see
Fig. \ref{fig:lf}) using our luminoisty
functions. This distribution is directly comparable with the
observations and model independent.
 Our calculation in each cosmology is based
on this quasar distribution. So the small cosmological dependence of the
 luminosity function is accounted for. The left panel also
indicates three redshift intervals ($z \sim 1, 2$ and 4), for which we show
the surface density of quasars as a function of magnitude limit. The
dashed lines show the corresponding results of \citet{Hen07b}. Our results are consistent with
theirs for the two low redshift intervals, but for the highest redshift
interval ($z \sim 4$), our quasar surface density is roughly a factor of
two of that of \citet{Hen07b} between $i=18$ and 22.
As we use the updated results by \citet{Richards05},
our luminosity function may be more realistic than that
in \citet{Hen07b}. It is worth emphasising, however, that this difference has only
moderate effects on the number of predicted multiply-imaged quasars
because most ($\sim 75\%$ and $\sim 70\%$) of the lensed sources are
below redshift 3 for the photometric and spectroscopic samples. 

\section{RESULTS}

\begin{figure}
{
 \centering
 \leavevmode 
\columnwidth=1.0\columnwidth
\includegraphics[width={\columnwidth}]{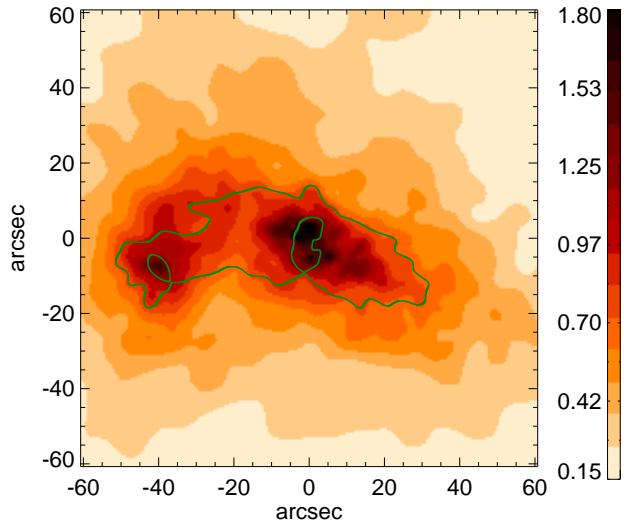}%
}
\caption{The surface density map (in units of the critical surface
density) for the most massive cluster at redshift $\zl=0.5$ in the
$\LCDM$0 simulation (in one projection).
The cluster virial mass is $1\times 10^{15}h^{-1}M_{\odot}$. The colour bar at
the right shows the scales. Notice that the
main cluster (located at the origin) is  merging with a sub-cluster
at the left. The critical curves (appropriate for a source at redshift 2.5)
are shown as the green curves (see also Fig. \ref{fig:caustics}.)}
\label{fig:cluster}
\end{figure}

\begin{figure}
{
 \centering
 \leavevmode 
\columnwidth=.5\columnwidth
\includegraphics[width={\columnwidth}]{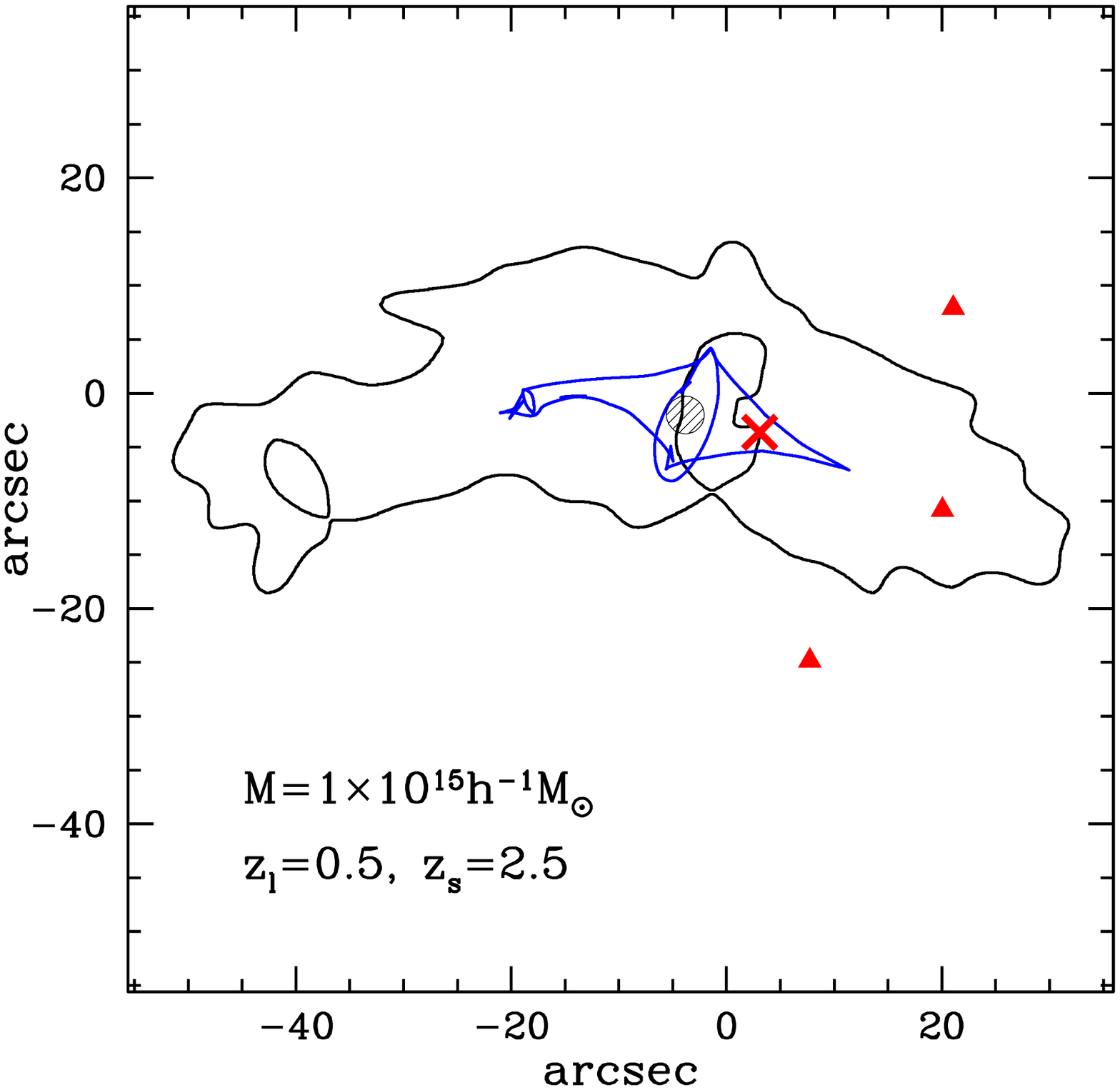}%
\includegraphics[width={\columnwidth}]{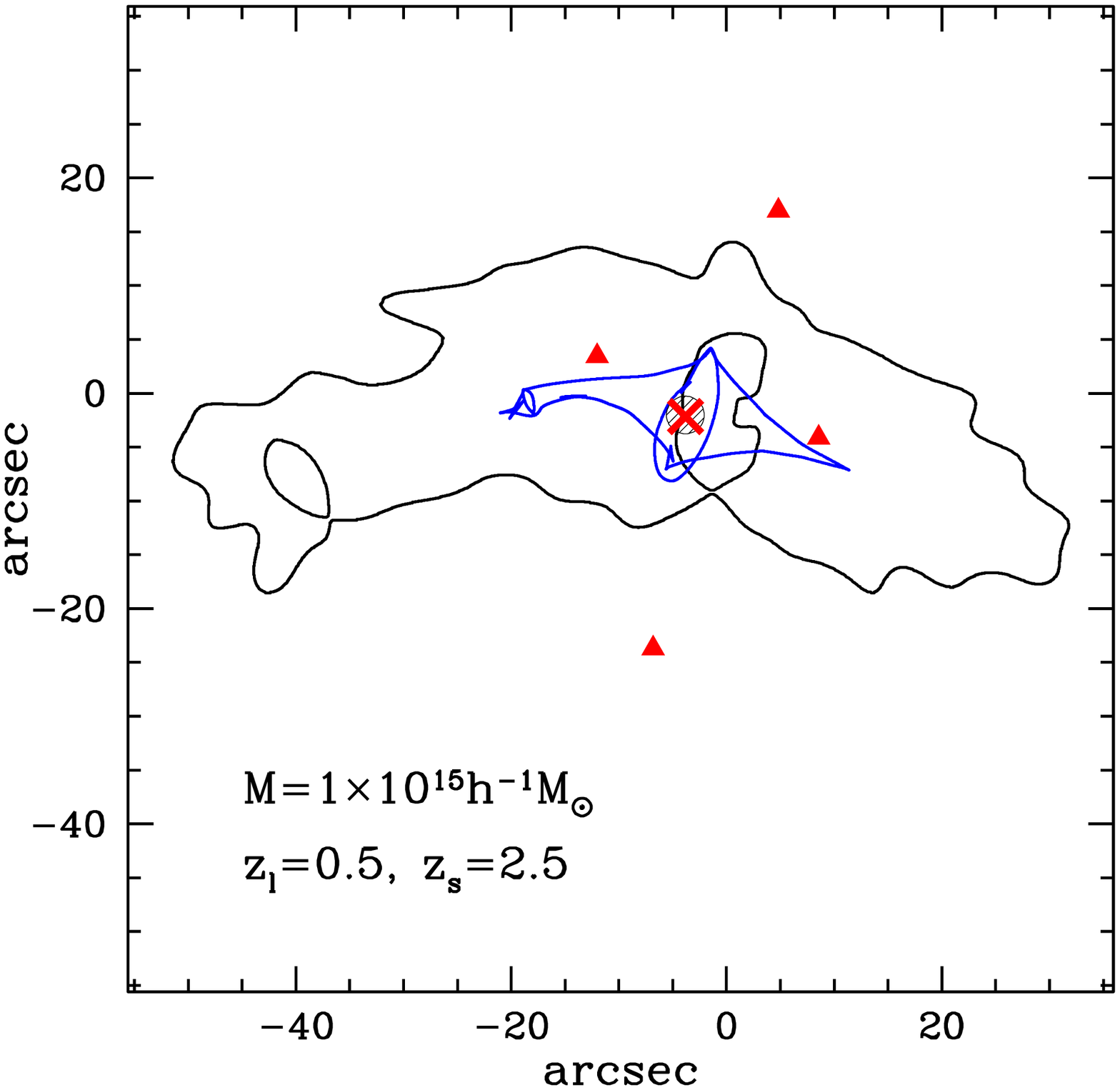}%

\includegraphics[width={\columnwidth}]{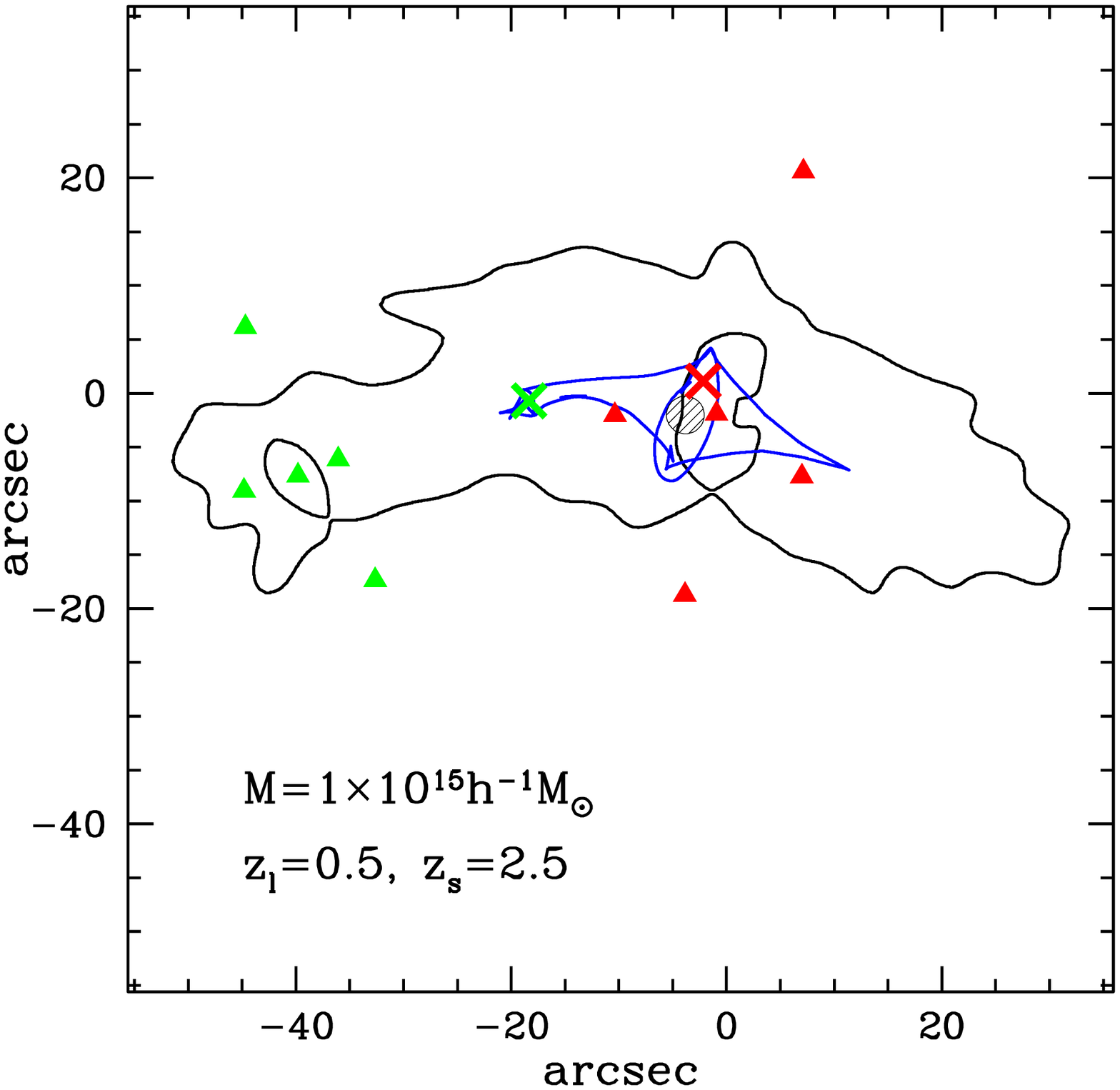}%
\includegraphics[width={\columnwidth}]{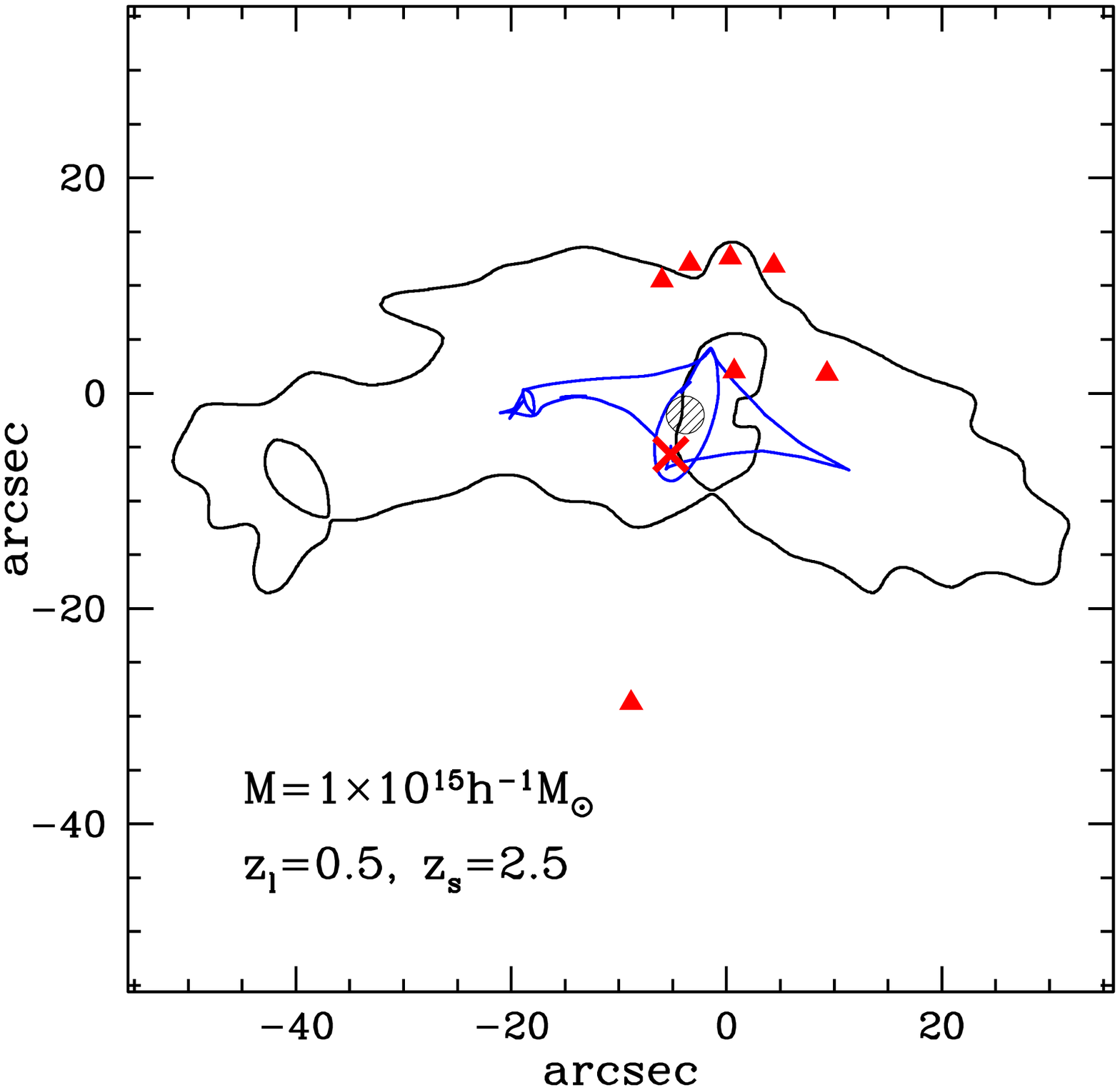}%
}
\caption{Example image configurations for the most massive
cluster at $\zl=0.5$ in our $\LCDM$0 simulation shown in Fig. \ref{fig:cluster}.
The source is assumed to be at redshift $\zs=2.5$. The cluster mass is $1\times 10^{15}h^{-1}
 M_{\odot}$. The blue and black curves are the caustics and critical
 curves respectively. For each panel, the red cross shows the
source position and the corresponding images are shown as triangles.
The bottom left panel shows an additional source position (green
cross) and the corresponding images are shown as green triangles.
When a source lies in the shaded circle, the central image 
disappears. This is a result of approximating the Brightest Cluster Galaxy
(BCG) as a truncated isothermal sphere (see the discussion in \S\ref{sec:images}).
}
\label{fig:caustics}
\end{figure}

Armed with knowledge of the background source population and cluster
lensing cross-sections, we can now make detailed predictions for
multiply-imaged quasars in the SDSS. We start with an example cluster
and illustrate possible multiply-imaged configurations in
\S\ref{sec:images}. We then discuss statistical properties 
of multiply-imaged quasars, including the total number and partitions in
different image multiplicities, for both the $\LCDM$0 and WMAP3
cosmologies in \S\ref{sec:number}. In \S\ref{sec:central} we examine
the properties of central images, and finish with a discussion
of cosmic variance in \S\ref{sec:cosmic_variance}.

\subsection{An example cluster: caustics, critical curves and images \label{sec:images}}

Fig. \ref{fig:cluster} shows the projected surface density map
(along one direction) for the most massive cluster in the $\LCDM$0
cosmology at redshift 0.5. This cluster has a mass of about $10^{15}h^{-1}M_\odot$.
It can be seen that in addition to the massive cluster
located close to the origin, there is another sub-cluster located
roughly $40\arcsec$ away (corresponding to $170h^{-1}\kpc$ away in
projection) to the left merging with the main cluster shown at the origin.

The critical curves and caustics for this cluster are shown in
Fig. \ref{fig:caustics}. Both are quite elongated along the horizontal
axis, partly due to the shear of the merging sub-cluster. Notice that there
is also a small, inner critical curve associated with the sub-cluster. 
The caustics
consist of an inner ellipse and an outer diamond (there are
small intersections, however). There are
also high-order singularities such as swallowtails in the caustics.
Sources located inside such high-order singularity regions can
produce image numbers higher than 5 (see the bottom right panel). 

The top left panel of Fig. \ref{fig:caustics}
shows an image configuration where the source is located just inside the
outer diamond caustic. All three formed images are quite far away from the
centre of the main cluster. When the source moves into the inner
ellipse, we normally expect five images. However, as a consequence of modelling
the BCG as a truncated singular isothermal sphere, we find that there is 
a roughly circular region (shaded in the Figure) within which the central, fifth image
disappears -- the central image has been `swallowed' by the BCG. The radius of this region
is roughly one angular Einstein radius corresponding to the
isothermal sphere, $4\pi (\sigma/c)^2 D_{\rm ls}/D_{\rm  s}$, 
where $D_{\rm ls}$ is the angular diameter distance between the lens and
source, and $\sigma$ is the velocity dispersion (cf. eq. \ref{eq:sigma}). Such a four-image configuration is illustrated in the top right
panel. We can regard such cases as an extreme example of a five-image configuration where the
central image has been infinitely de-magnified. 

The bottom left panel in Fig. \ref{fig:caustics} 
shows a source position that produces the usual five-image systems. As expected,
the fifth image is close to the cluster
centre. There is, however, a different kind of five image
configuration, which is produced by the 
presence of the sub-cluster. As we only put in (by hand) a BCG at the 
centre of the main cluster, the central image induced by such 
sub-clusters is always present, and will be located close to the centre
of the sub-cluster, and in general much further away from the main
cluster centre (see also Fig. \ref{fig:nimage5}). 
%This will be an approximation if the sub-clusters also sink in together with a BCG.

For rare cases, when the source is located in regions of high-order
singularity, the image number can exceed 5. One example is shown in the
bottom right panel. For this case, four images are clustered close
together, tangentially aligned with a critical curve,
while the other three images are further apart. It can be seen
from Fig. \ref{fig:caustics} that such high-order singularity regions are
rare, and so we do not expect a significant fraction of sources to have
image numbers exceeding five.

For other clusters in our simulations, the image configurations are similar
except that for some small clusters, we can also have two-image
configurations. These are the usual three-image systems for which the
central image has been `swallowed' by the BCG, similar to the four-image
case discussed above.

\subsection{Expected number of multiply-imaged quasars in the SDSS \label{sec:number}}

\begin{figure}
{
 \centering
 \leavevmode 
\columnwidth=.95\columnwidth
\includegraphics[width={\columnwidth}]{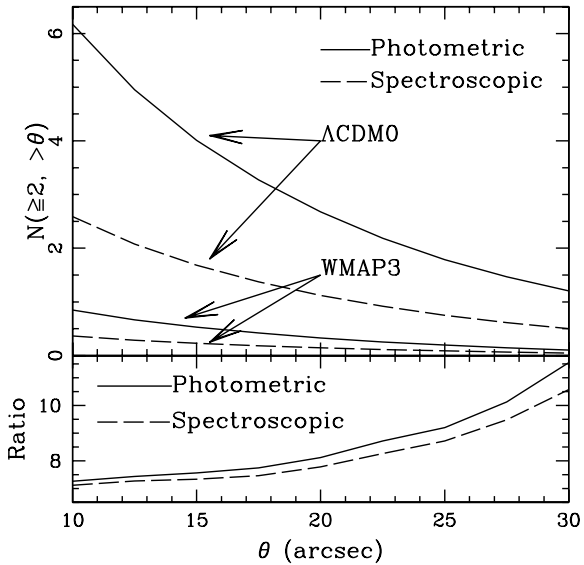}%
}
\caption{The top panel shows the number of predicted 
large-separation multiply-imaged quasars as
function of separation for 
the $\LCDM$0 (top two curves)
and WMAP3 (bottom curves) models, while the solid 
and dashed lines are for the SDSS photometric 
 (with an effective area of $\sim$ 8000 deg$^2$) and
 spectroscopic (with an effective area of $\sim$ 5000 deg$^2$) quasars
  sample respectively.
The bottom panel shows the ratio of the predicted numbers in the two
cosmologies. Notice that the ratio increases as the separation increases.
} 
\label{fig:number}
\end{figure}

The expected number of multiply-imaged quasars in the SDSS is shown
in Fig. \ref{fig:number} for the photometric (solid) and
spectroscopic (dashed) samples, in both the $\LCDM$0 (top two curves) and
WMAP3 (bottom curves) cosmologies. In the $\LCDM$0 cosmology, we expect
6.2 and 2.6 lenses with separations larger than $10\arcsec$ for the
photometric (with an effective area of $\sim$ 8000 deg$^2$) and
 spectroscopic (with an effective area of $\sim$ 5000 deg$^2$) 
samples respectively. 
The corresponding numbers are much lower in 
the WMAP3 model, about 0.85 and 0.36 respectively.
The bottom panel of Fig. \ref{fig:number} shows the relative ratios in
these two cosmologies -- the ratio is about
$\sim 7$ at $10\arcsec$, but increases to about $\sim 11$ at
$30\arcsec$, for both samples.

The large reduction factor in the WMAP3 model is a direct result of the lower
abundance of clusters in this cosmology compared with $\LCDM$0
(see Fig. \ref{fig:variance}) because of the lower
$\sigma_8$ and lower mass density $\omega0$. The reduction factor 
is larger for more massive clusters. This explains the separation dependence 
of the ratio because larger separation lenses are produced by more massive clusters.
The reduction factor is comparable to that of giant arcs 
($\sim 6$) for $\theta> 10\arcsec$ but larger by a factor of 2 for  $\theta> 30\arcsec$.
The substantially lower number of predicted multiply-images in the WMAP3 model may be
somewhat difficult to reconcile with the data; we will return to this point at the end of this section.

\begin{figure}
{
 \centering
 \leavevmode 
\columnwidth=.95\columnwidth
\includegraphics[width={\columnwidth}]{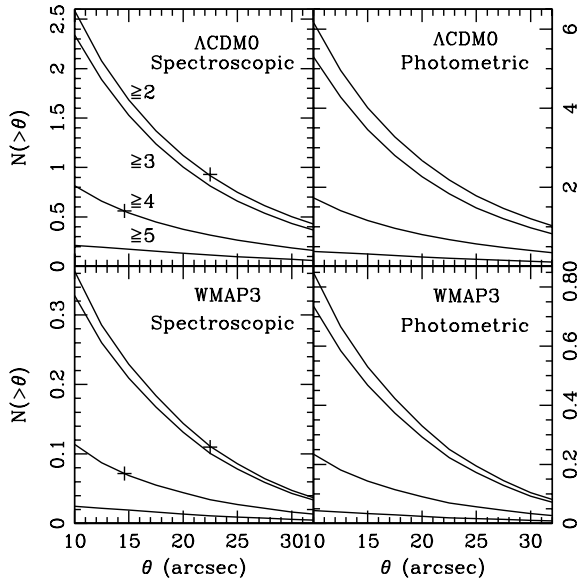}%
}
\caption{The number of large-separation multiply-imaged quasars
as function of separation and image multiplicity for the SDSS
photometric (with an effective area of $\sim$ 8000 deg$^2$) and
 spectroscopic (with an effective area of $\sim$ 5000 deg$^2$)
 quasar samples. The top and bottom panels
show the results for the $\LCDM$0 and WMAP3 cosmologies respectively.
%The cosmology and quasar sample are labeled at the top right.
For each panel, from top to bottom, the lines are for $\Nimage \ge 2, 3, 4, 5$
respectively. The two crosses show the predicted positions of the two known large-separation lenses.
} 
\label{fig:multi}
\end{figure}

We can further divide the predicted number of multiply-imaged quasars
by clusters according to the image multiplicity. The results are shown
in Fig. \ref{fig:multi} for the two quasar samples in the $\LCDM$0
and WMAP3 cosmologies. While the amplitudes differ, the relative ratios
of lenses with $\Nimage\ge$ 2, 3, 4, and 5 are quite similar.
For example, the number of systems with $\Nimage \ge 4$ is about 
1/3.5 of the number of systems with $\Nimage \ge 2$. 
%The number of
%systems with more than 5 images is even smaller, only about 10\% of the total
%number of lenses.

\begin{figure}
{
 \centering
 \leavevmode 
\columnwidth=0.5\columnwidth
\includegraphics[width={\columnwidth}]{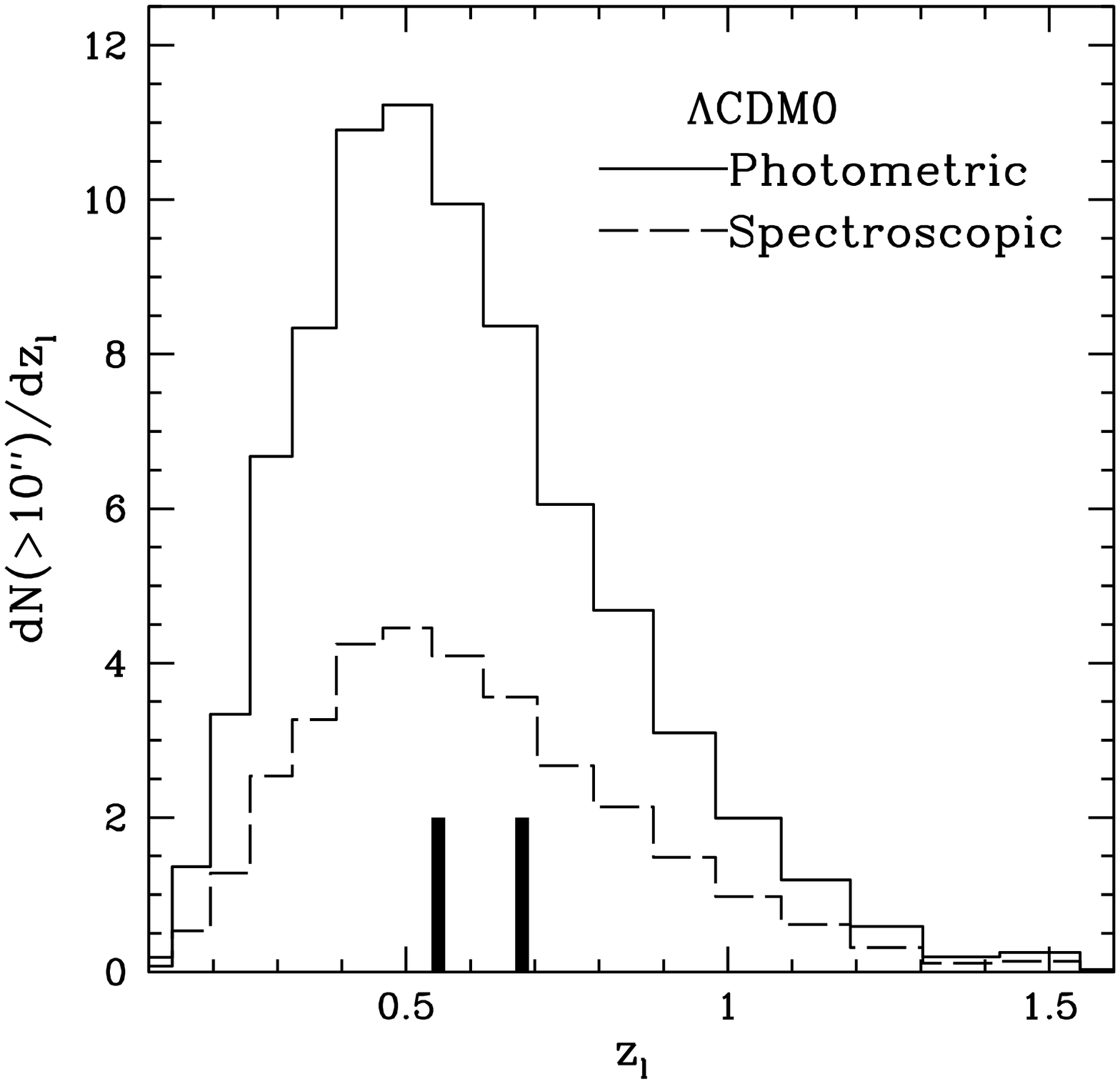}%
\includegraphics[width={\columnwidth}]{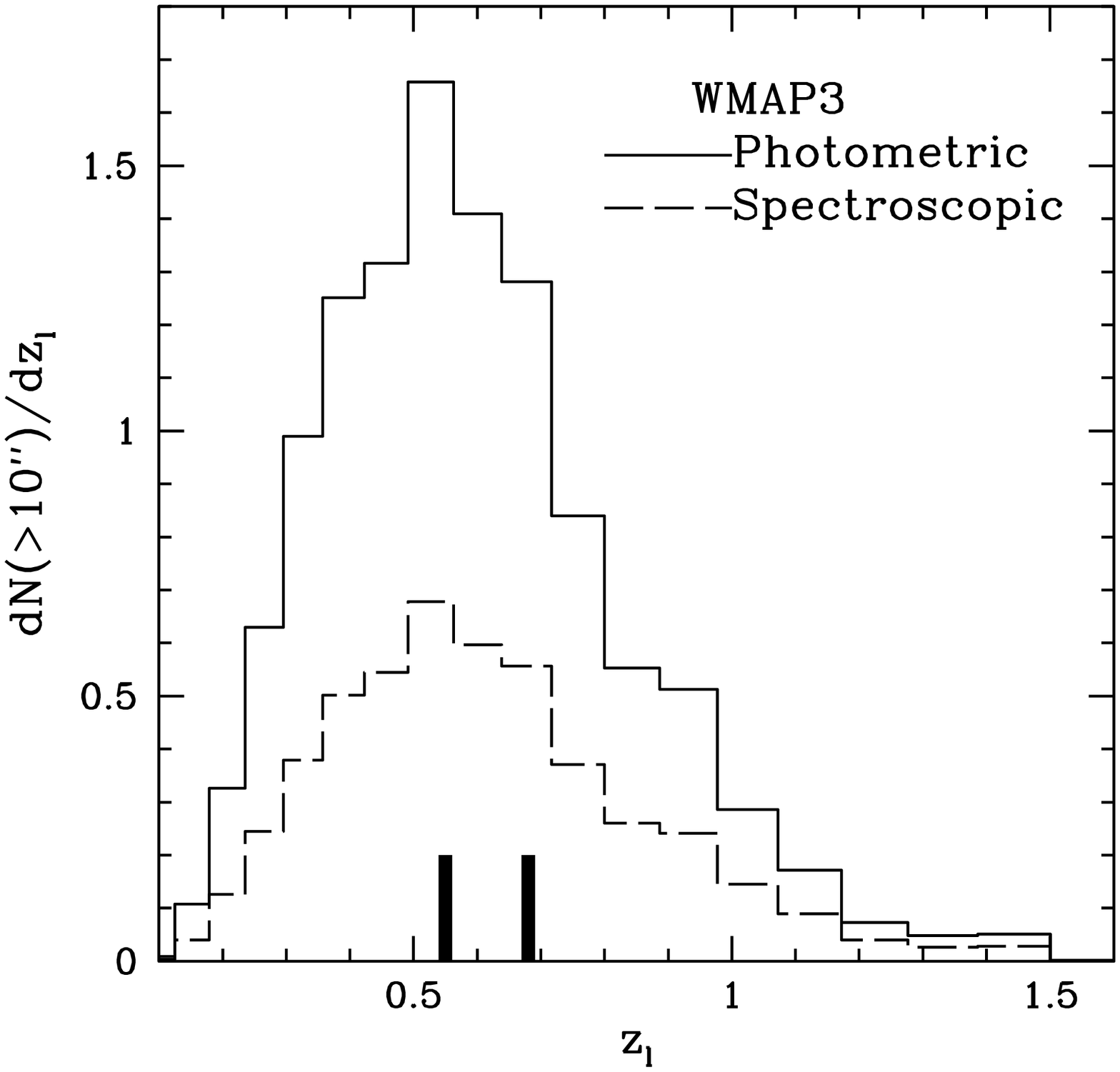}%
}
\caption{The lens redshift distribution in the $\LCDM$0
  cosmology (left) and WMAP3 cosmology (right). The area under each
  curve is normalised to the total number of multiply-imaged quasars.
  The two vertical bars show the lens redshifts of SDSS J1004+4112 (0.68) and
SDSS J1029+2623 (0.55). The shapes are similar in these two cosmologies. 
Notice that the median lens redshift is slightly lower for the photometric quasar
  sample (see \S\ref{sec:lf}).
} 
\label{fig:lens}
\end{figure}

It is interesting to ask for multiply-imaged quasars, what are the
typical redshifts of lensing clusters and background quasars. These are
important for observationally identifying the lensing
candidates. Fig. \ref{fig:lens} shows the differential number
distribution for the lens redshift. The distributions are similar
in the $\LCDM$0 and WMAP3 cosmologies; both peak around 
redshift 0.5, and 90\% of the lensing clusters are predicted  to be
between redshift 0.2 and 1. The predicted lens redshift
distribution is in good agreement with \citet{Hen07b}.
Notice that the median redshift for the 
lenses in the spectroscopic sample is slightly higher than that
for the photometric sample. This can be understood as follows:
quasars in the spectroscopic sample are on average at higher redshift, 
since the most likely lens location is roughly mid-way between the lens and
source, as a result the lens redshift is also slightly higher.
The two observed cluster lenses have redshifts 0.55 and 0.68, roughly at
the position of the peak, in good agreement with the theoretical expectations.

\begin{figure}
{
 \centering
 \leavevmode 
\columnwidth=.5\columnwidth
\includegraphics[width={\columnwidth}]{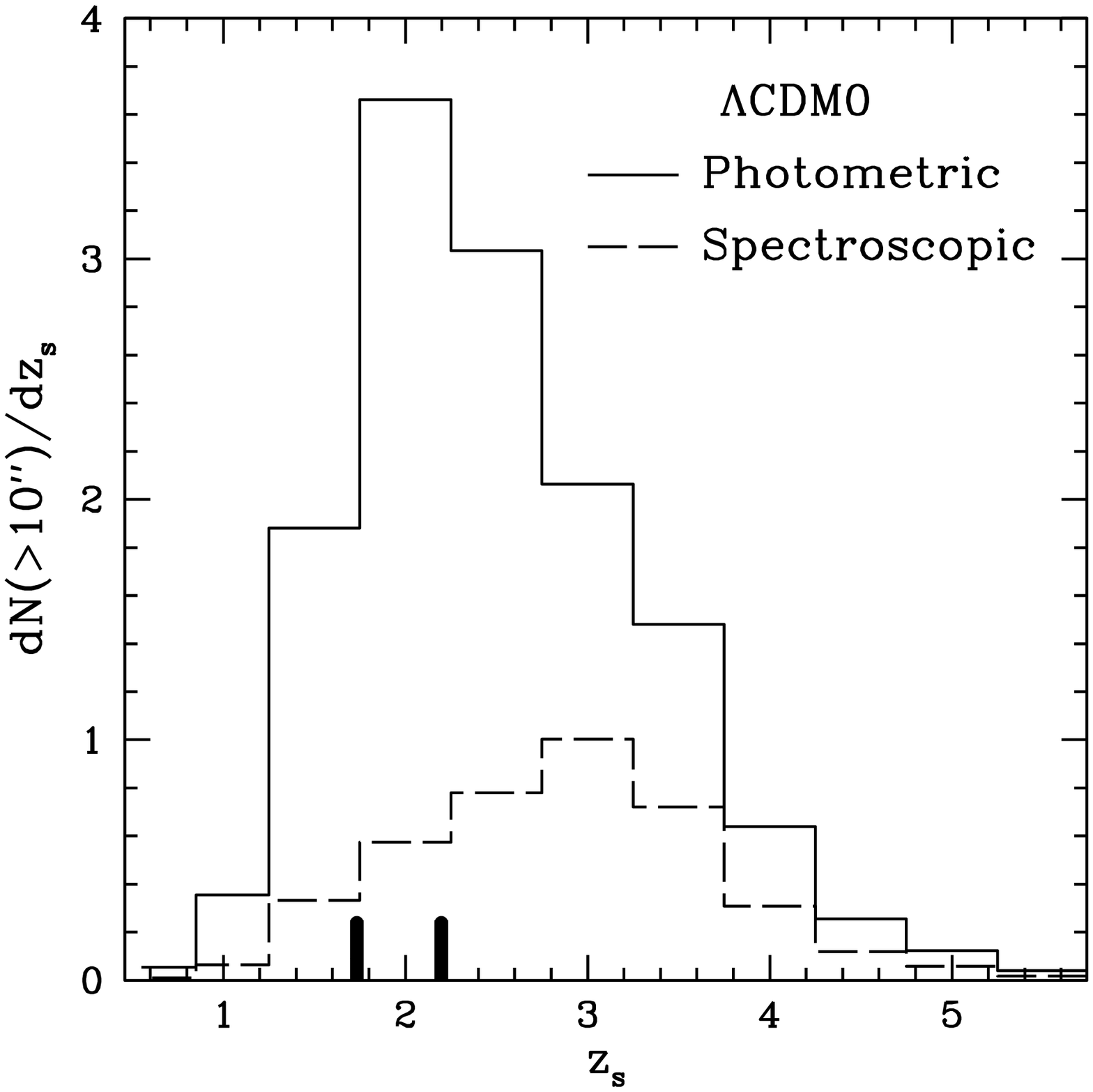}%
\includegraphics[width={\columnwidth}]{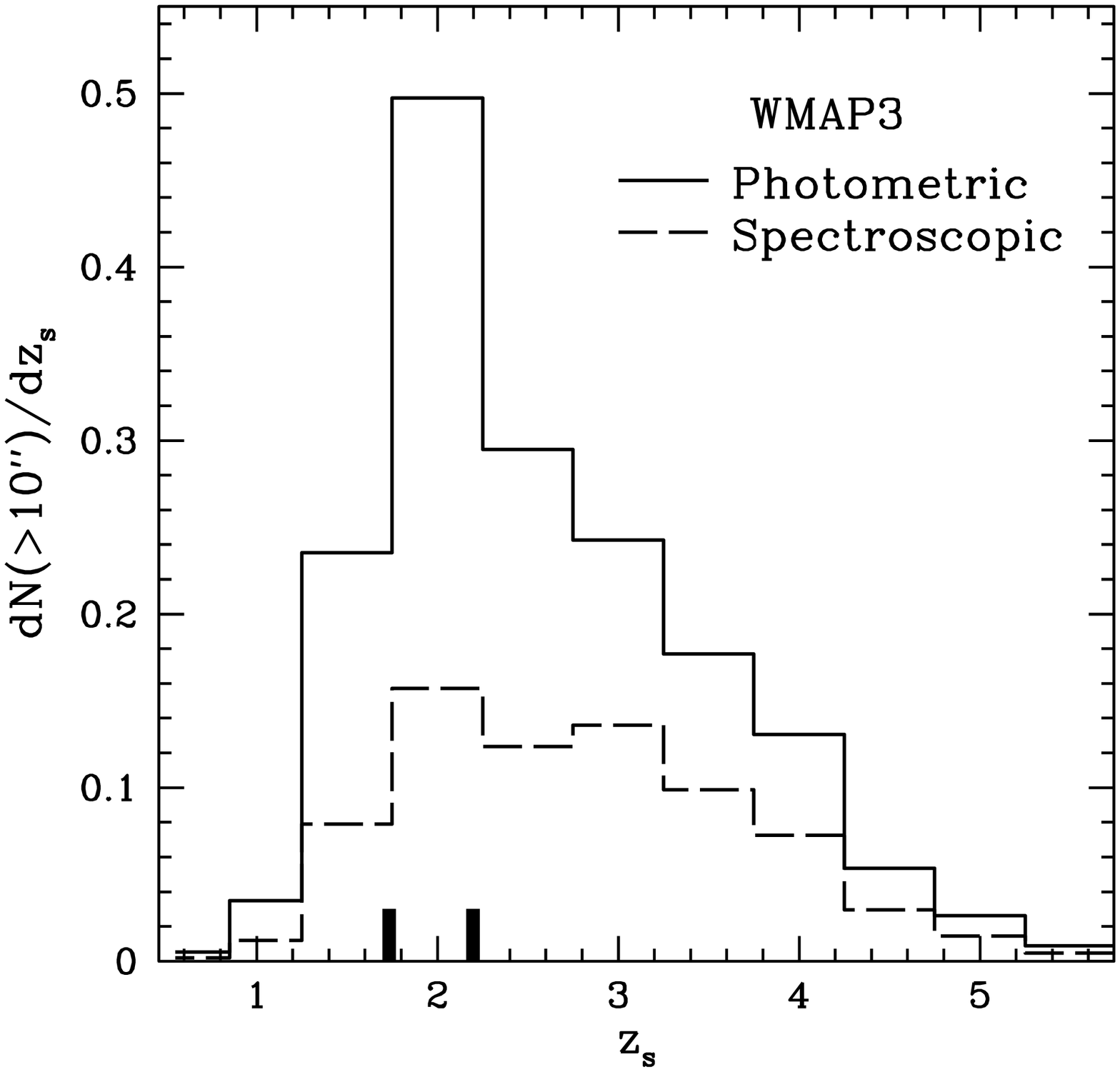}%
}
\caption{The source redshift distribution of multiply-imaged quasars
 with separation $>10\arcsec$ 
in the $\LCDM$0 (left) and WMAP3 (right) models. The peak of the
source redshift distribution shifts to lower redshift in the photometric
sample due to the lower mean quasar redshift for this sample (see \S\ref{sec:lf}).
The vertical bars show the sources redshifts for SDSS J1004+4112 (1.734) and
SDSS J1029+2623 (2.197).
} 
\label{fig:source}
\end{figure}

The corresponding source redshift distribution is shown in
Fig. \ref{fig:source}. The distributions are again similar in the
two cosmologies. For the multiply-imaged quasars in the photometric
sample, their redshift distribution peaks around redshift 2, while for the spectroscopic quasar 
sample, the peak is extended with a range from 2 to 3. 
This is again a result of the
lower source redshift in the photometric sample (see \S\ref{sec:lf}).
The two observed source redshifts (1.734 for J1004+4112, and 2.197 for
J1029+2623) coincides with the peak predicted for the photometric
sample, but appears to be at the low side for the the spectroscopic
sample (but unlikely to be statistically significant due to the limited
number statistics).

It is interesting to explore whether these two cosmologies are consistent
with the data, given even the very limited number of cluster lenses.
The effective area for the Data Release 5 of the SDSS is 5740 deg$^2$.
J1029+2623 is from SDSS-II, which has now discovered about 10,000 QSOs, which 
is roughly 1/4 of the QSOs discovered in the Data Release 3, which has
an area of about 3700 deg$^2$, so these two lenses were discovered
in an effective area of $\sim 5740+3700/4 \approx 6500$ deg$^2$ in the SDSS spectroscopic sample.
In the $\LCDM$0 model, the two crosses in the top left panel in Fig. \ref{fig:multi} shows the positions we 
predict, (14.6, 0.55) and (22.5, 0.93) for these two systems. 
The number of lenses with separation larger than
$14.6\arcsec$ quadrupole-images (J1004+4112) and $22.5\arcsec$ double images (J1029+2623) are
$0.55\times6500/5000=0.7$ and $0.93\times6500/5000=1.2$. 
This is roughly consistent with what we have discovered in the SDSS so far. 
However, for the WMAP3 model, the positions of crosses in the bottom left panel 
in Fig. \ref{fig:multi} are (14.6 0.072) and (22.5 0.094).
The number of lenses with separation larger than
$14.6\arcsec$  quadrupole-images and $22.5\arcsec$ double images are
$0.072\times6500/5000=0.094$ and $0.11\times6500/5000=0.14$ for the SDSS 
spectroscopic sample with an effective area of $\sim$6500 deg$^2$. 
We conservatively estimate the probability of observing such two 
systems in WMAP3 model as follows. The predicted number of lensed quasar
systems with $\Nimage\geq 2$ and $\theta> 10\arcsec$ is 0.36$\times$6500/5000=
 0.468. Assuming a Poisson distribution with a mean number of lenses $\lambda=0.468$,
the probability of observing these two systems in WMAP3 model is,
 $P(N\geq2)=1-P(0)-P(1)=1-e^{-\lambda}-\lambda e^{-\lambda}\sim8\%$.
So even with the very limited
sample of two multiply-imaged quasars, it seems that
the WMAP3 cosmology is compatible with observations only at $\sim
8\%$ level.
 Considering that our lensing optical depth may be over-estimated
in the WMAP3 model due to cosmic variance (see \S\ref{sec:cosmic_variance}),
 and the fact that the number of observed systems may be
a lower limit (due to possible incompleteness in the surveys), so
 the probability may be even lower ($\lesssim 4\%$) in reality.

\subsection{Properties of central images \label{sec:central}}

For non-singular lensing potentials, catastrophe theory predicts
generically an odd number of images (\citealt{Burke81}). However,
for nearly all galaxy-scale lenses
(with the exception of PMN J1632-0033, \citealt{Winn03, Winn04}),
the number of images detected is even. It is
thought that the central images may have been highly demagnified by
the stars at the centre of lensing galaxies (e.g., \citealt{Na86, Keeton03}).
Furthermore the central images can be 
swallowed by the presence of a central black hole (e.g., \citealt{Mao01,
Bowman04, Rusin05}). Even when a central image is present,
it may not be easily identifiable because the central
cluster galaxy is usually very bright. In this regard, colour information
may be particularly useful in separating the (usually bluer) central QSO image
from the central cluster galaxy.

Extrapolating our experience for galaxy-scale
lenses, we expect the central images on cluster-scale lenses will provide 
strong constraints on their central mass profiles.
It is therefore interesting to explore the properties of central images in
cluster lenses. Below we discuss three image and five image cases in
turn. Here, we take all of the images in the lens systems into account 
regardless how faint they are. 
For definiteness, we choose a cluster population at $\zl=0.5$ and a
source population at $\zs=2.5$, roughly at the peak of the lens and source redshift distributions.

\begin{figure}
{
 \centering
 \leavevmode 
\columnwidth=.5\columnwidth
\includegraphics[width={\columnwidth}]{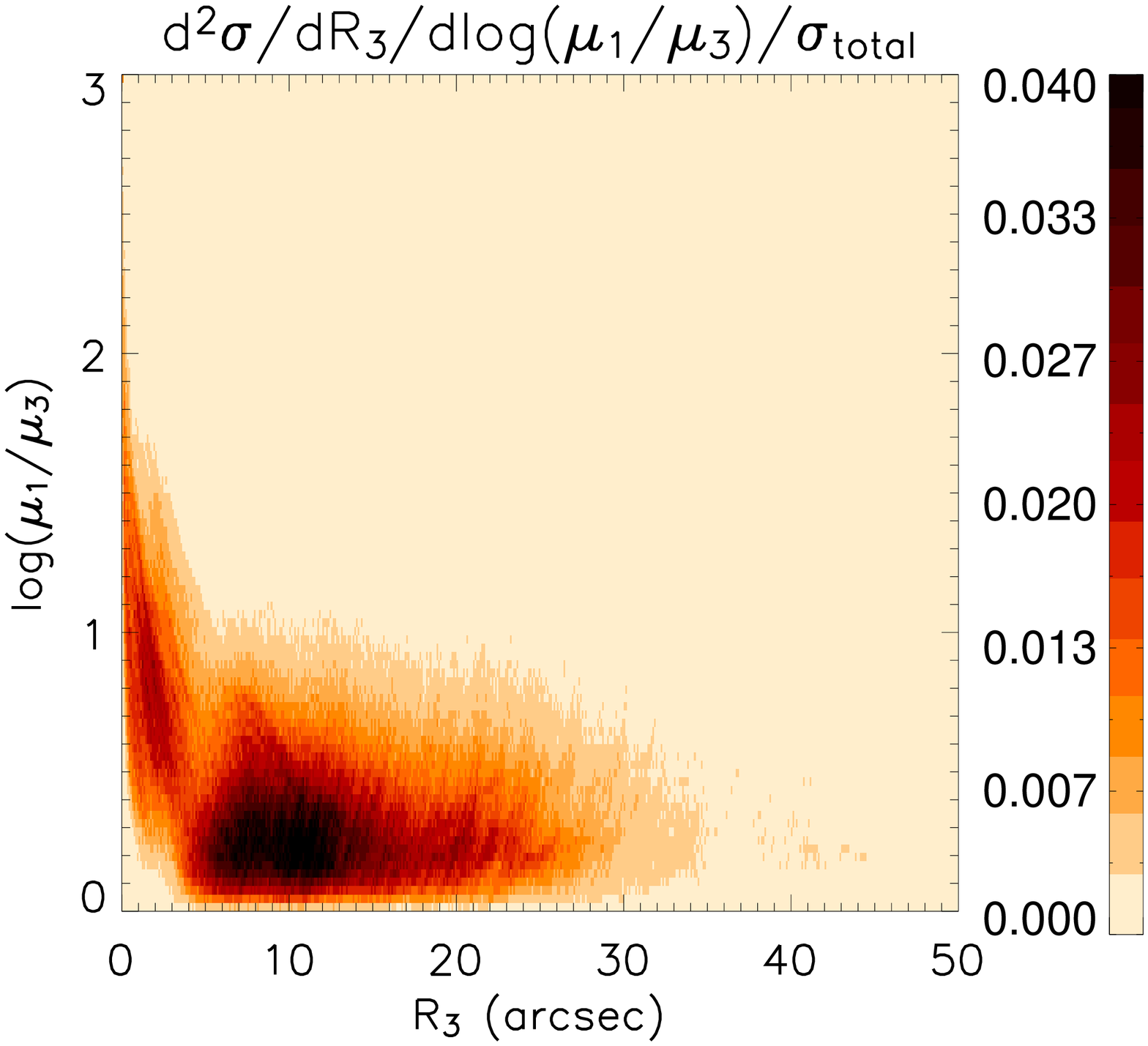}%

\includegraphics[width={\columnwidth}]{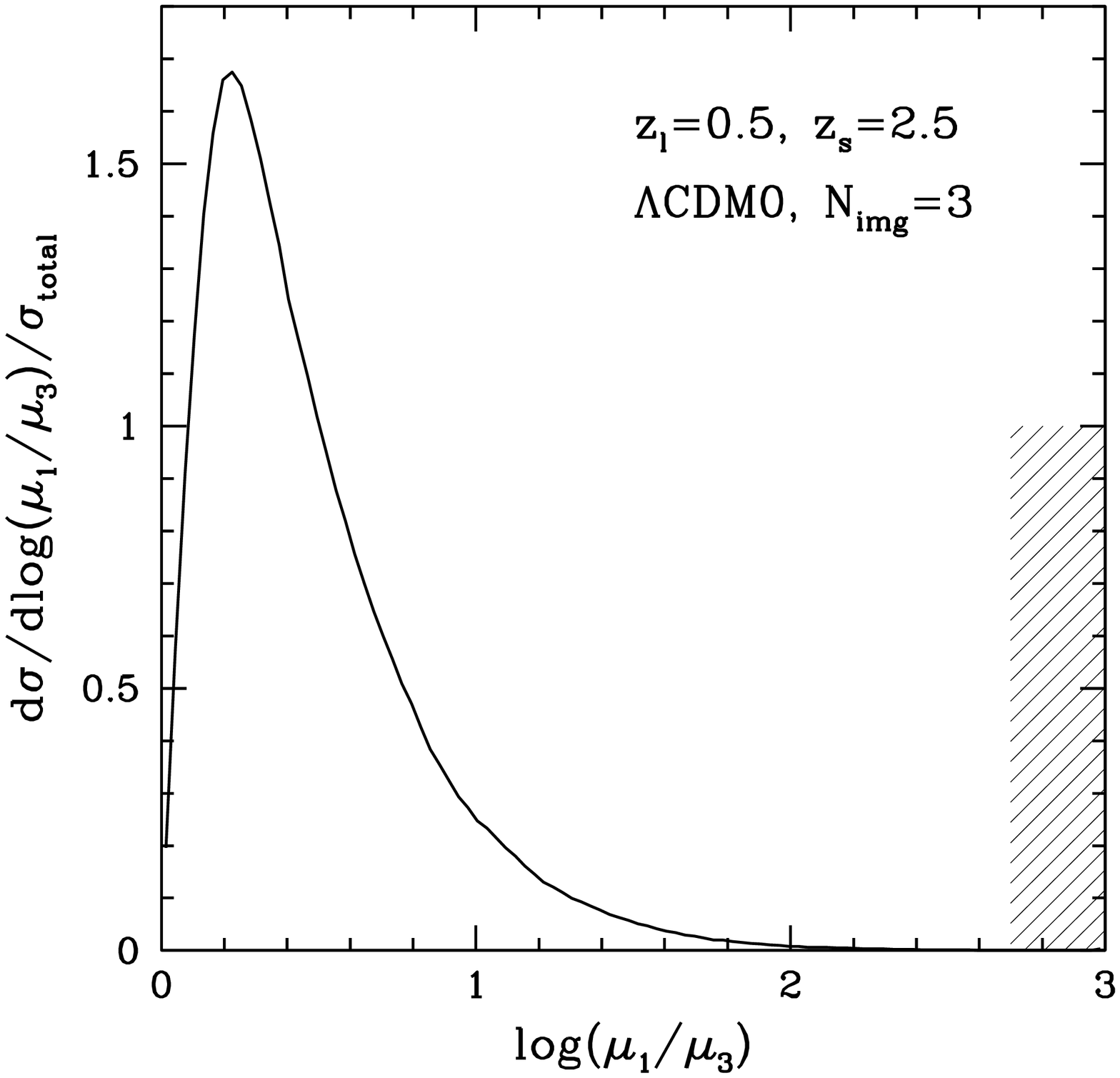}%
\includegraphics[width={\columnwidth}]{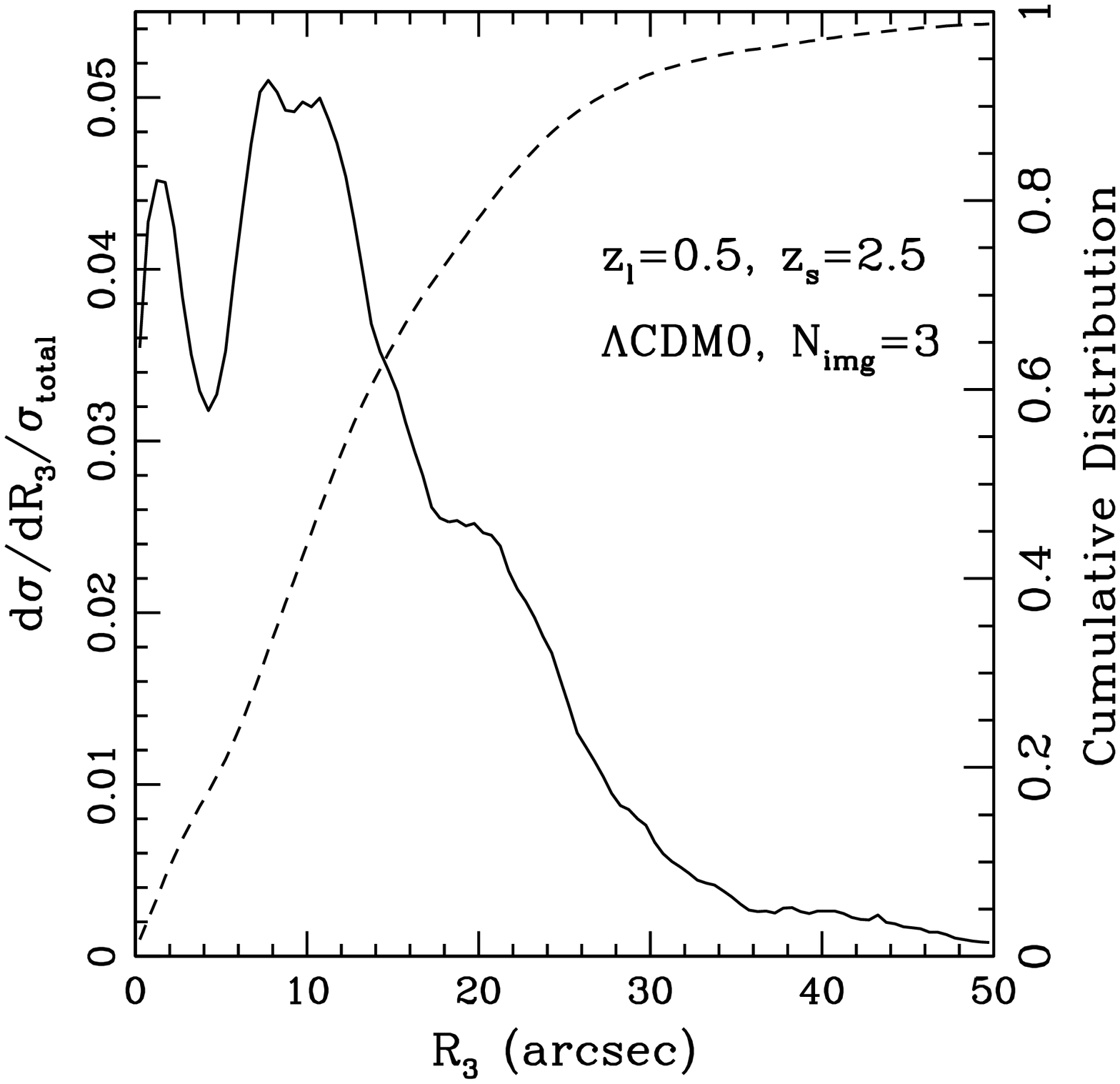}%
}
\caption{The top shows the normalised differential cross-section
 for three-image systems in the parameter space of
 $\log \mu_1/\mu_3$ and $R_3$ for the $\LCDM$0 cosmology, here 
 $R_3$ is the separation of the central image relative to the cluster
 centre, $\mu_1$ and $\mu_3$ are  magnifications for the brightest and
 faintest images; the latter is taken as the central image. The cluster 
is at $\zl=0.5$ and the source is at $\zs=2.5$. The bottom left panel
 shows the differential probability distribution of $\log
 \mu_1/\mu_3$. The hatched region shows the corresponding cross-section
 for two-image systems, i.e., three-image systems with an infinitely
 de-magnified central image due to the BCG. The bottom right panel
 shows the probability distribution of $R_3$. The cumulative
 distribution is shown as the dashed line.
}
\label{fig:nimage3}
\end{figure}

The top panel of Fig. \ref{fig:nimage3} shows the distribution of the differential cross-section 
of three-image systems in the plane of $\log \mu_1/\mu_3$ and $R_3$, here
$\mu_3$ and $\mu_1$ are the magnification of the faintest and brightest
image, and $R_3$ is the distance of the faintest image from the cluster
centre. This plot shows two distinct, and disjoint regions.
The bottom region arises due to three-image configurations illustrated 
in the top left panel of Fig. \ref{fig:caustics}; in such cases, all
three images are magnified and have similar intensity ratios. The region
at much smaller $R_3$ are due to three-image systems with a
genuine central image close to the main cluster centre. 

In the bottom left panel of Fig. \ref{fig:nimage3} we show the probability distribution of $\log
\mu_1/\mu_3$. It can be seen most ($\sim 90\%$) of the three-image systems should have a
central image brighter than 10\% of the brightest images, and so should
be relatively straightforward to detect. This conclusion, however,
ignores the two-image systems for which the central image has been
infinitely demagnified. Such systems are shown as the hatched region, and
is about 1/3 of the total number of three-image systems. If we take such systems 
as extreme cases of three-image systems with $\mu_1/\mu_3\rightarrow \infty$, 
the fraction will decrease to 0.9/(1+1/3)=65\%. The probability
distribution of $R_3$ shows two prominent peaks (the bottom right
 panel of Fig. \ref{fig:nimage3}), one around $1\arcsec$
and the other around $10\arcsec$, followed by an extended tail. The peak
around $1\arcsec$ is again due to genuine third, central images, while
the peak around $10\arcsec$ and the extended tail are due to systems as
illustrated in the top left panel of Fig. \ref{fig:caustics}. Our
theoretical model appears to show for the two-image system, SDSS J1029+2623, 
the central image has a substantial chance of being observable given 
sufficiently deep images.

\begin{figure}
{
 \centering
 \leavevmode 
\columnwidth=.5\columnwidth
\includegraphics[width={\columnwidth}]{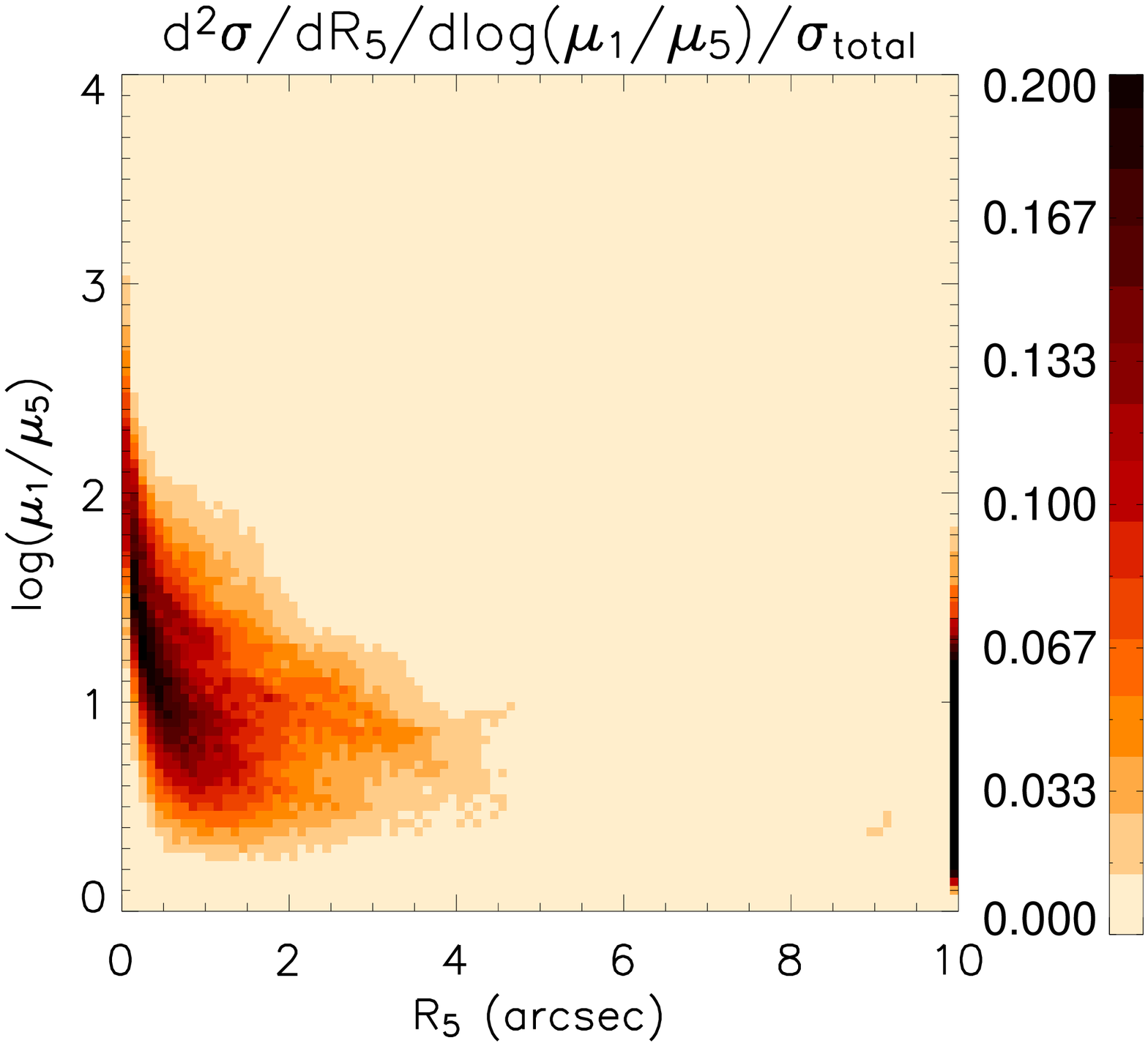}%

\includegraphics[width={\columnwidth}]{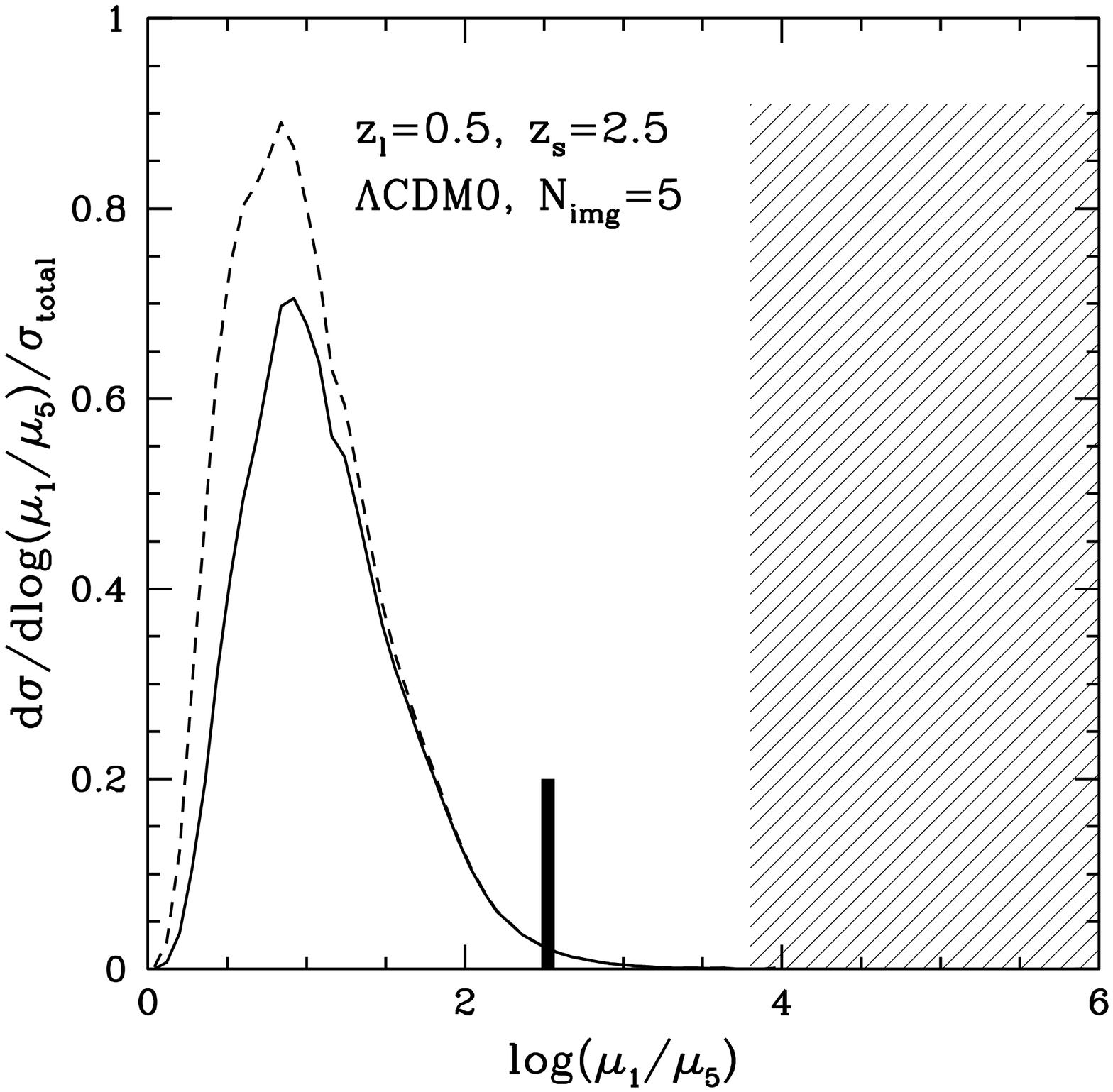}%
\includegraphics[width={\columnwidth}]{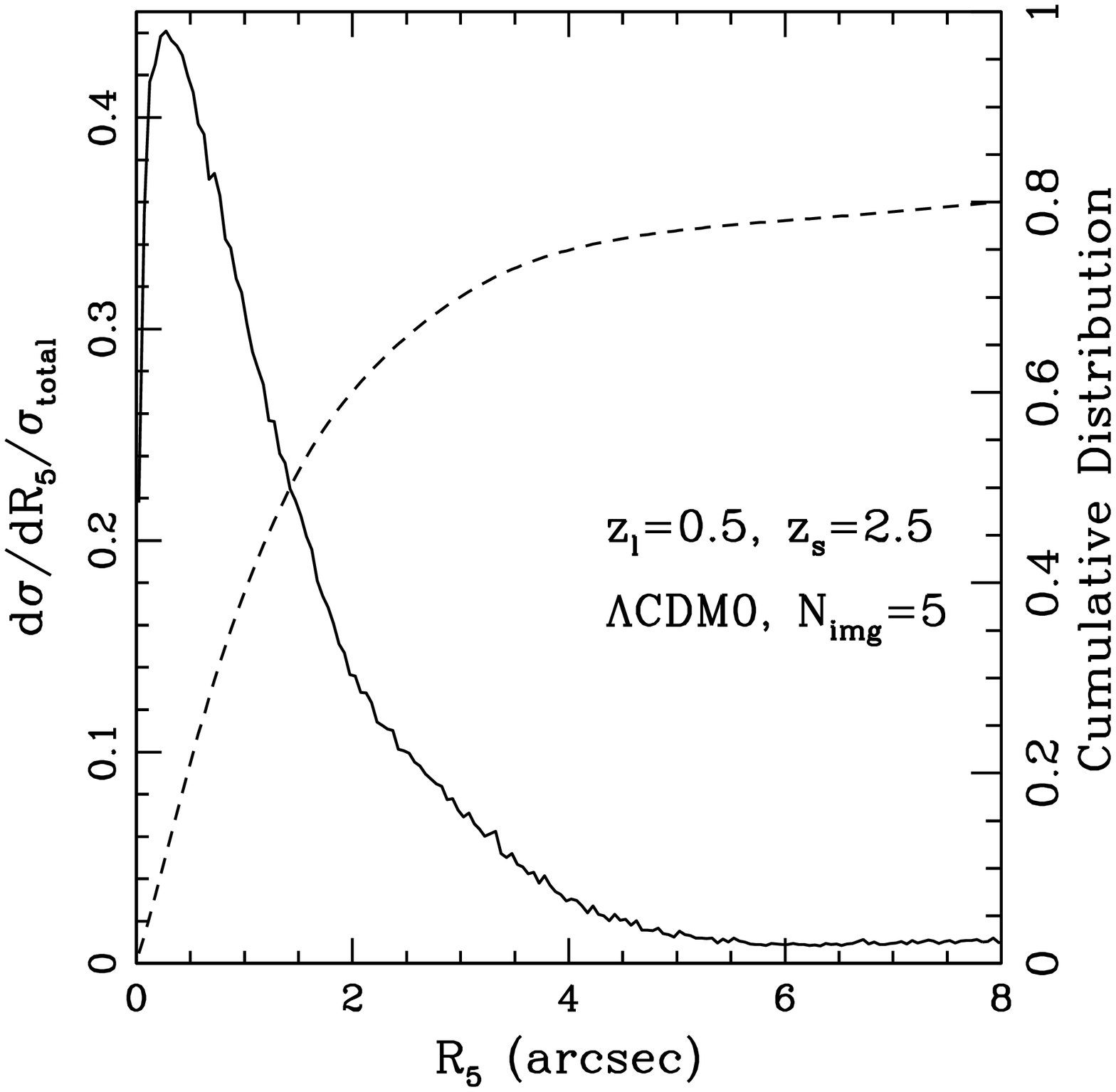}%
}
\caption{The top panel shows the normalised differential cross-section
 for five-image systems in the parameter space of
 $\log\mu_1/\mu_5$ and $R_5$ for the $\LCDM$0 cosmology, here 
 $R_5$ is the separation of the central image relative to the cluster
 centre, and $\mu_1$ and $\mu_5$ are the magnifications of
 the brightest and faintest images respectively, the latter is taken 
 as the central image. Systems with $R_5 >10\arcsec$ have been collected
 at the right edge pixels. The bottom left panel shows the probability
 distribution of $\log \mu_1/\mu_5$, which is simply obtained by integrating
 the left panel along the $x$-axis. The integration ranges are from 0
 to $5\arcsec$ for the dashed lines and from 0 to $\infty$ for the solid
 lines. The hatched region shows the corresponding cross-section
 for four-image systems, i.e., five-image systems with an infinitely
 de-magnified central image due to the BCG. The bottom right panel
 shows the probability distribution of $R_5$. The cumulative
 distribution is shown as the dashed line.
  }
\label{fig:nimage5}
\end{figure}

Of particular interests are the central images in systems with at least
four images, because for the first cluster quasar lens, J1004+4112,
subsequent deep imaging reveals a faint central image, 
at about 0.3\% of the brightest image in the system (see Table 1 in
\citealt{Inada05}). The top panel of  Fig. \ref{fig:nimage5} shows 
the number of five-image systems in the plane of $\log \mu_1/\mu_5$, where
$\mu_5$ is the magnification of the faintest image, and
$R_5$ is the distance of this image from the cluster centre. There are
also five-image systems with $R_5>10\arcsec$; they have been plotted
in the right-edge pixels in the left panel.

If we marginalise the separation from the cluster centre, we obtain the
probability distribution of magnification ratio, $\mu_1/\mu_5$,
shown in the bottom left panel of Fig. \ref{fig:nimage5}. The solid
line includes only those five-image systems with $R_5<5\arcsec$ while
the dashed line includes all five-image systems. The solid line excludes most
systems produced by merging sub-clusters. 
From these curves alone it seems that there is only a small probability of obtaining the
magnification ratio of 300, the observed value for J1004+4112. In fact, only 3\% will 
have a central image be fainter than the brightest one by a factor of 100.
However, this discounts those four-image systems where the central image has disappeared.
The hatched region is for those four image systems where the central image has been
infinitely de-magnified by the BCG. The total cross-section for
such systems is in fact quite large, about a factor of two of that for five-image systems.
 If we regard all these
four-image systems as five-image systems with infinitely demagnified central images,
the fraction will increase to $\sim (0.03+2)/(1+2)\sim68\%$.
If the profile is slightly shallower than the isothermal slope, $\rho \propto
r^{-2}$, these central images will have a very small but finite
magnification. Such central images may be what we are seeing in J1004+4112.
Clearly the brightness of the central image will be a very
sensitive probe of the central density profiles of BCGs.

The bottom right panel of Fig. \ref{fig:nimage5} shows the distribution of $R_5$. 
It can be seen that, roughly 50 per cent of five-image
systems have separation from the cluster centre smaller than $1.5\arcsec$.
About 25\% have separation larger than $5\arcsec$, these are due to
five-image systems produced by merging sub-clusters. If we include
four-image systems, and assume all the infinitely demagnified central images
have $R_5<1.5\arcsec$, then the vast majority
($\sim (0.5+2)/(1+2)\sim80\%$) of these systems will have central images within
$1.5\arcsec$ of the BCG. These properties are broadly consistent with the observed system
J1004+4112.

\subsection{Cosmic variance \label{sec:cosmic_variance}}

\begin{figure}
{
 \centering
 \leavevmode 
\columnwidth=.5\columnwidth
\includegraphics[width={\columnwidth}]{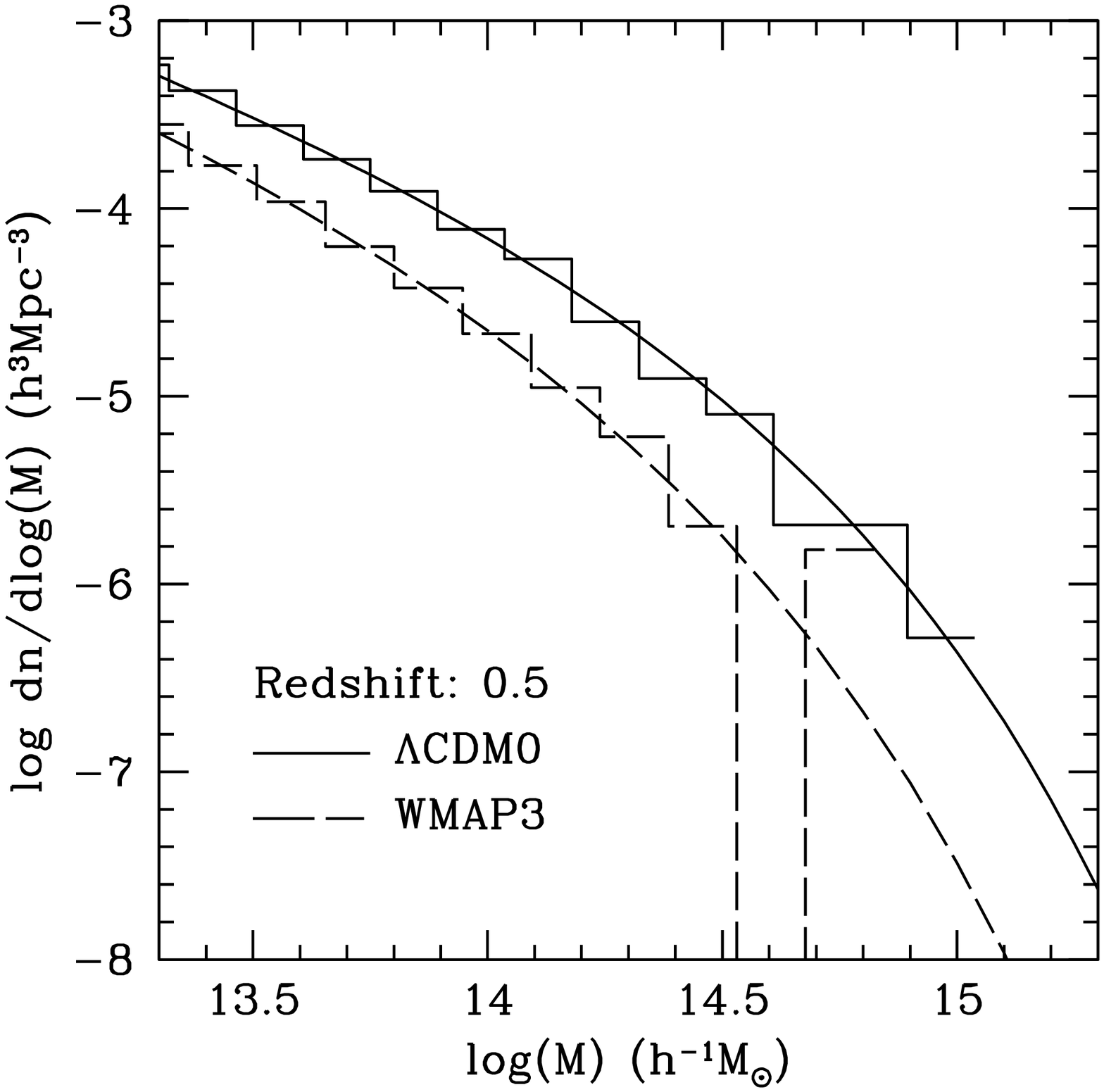}%

\includegraphics[width={\columnwidth}]{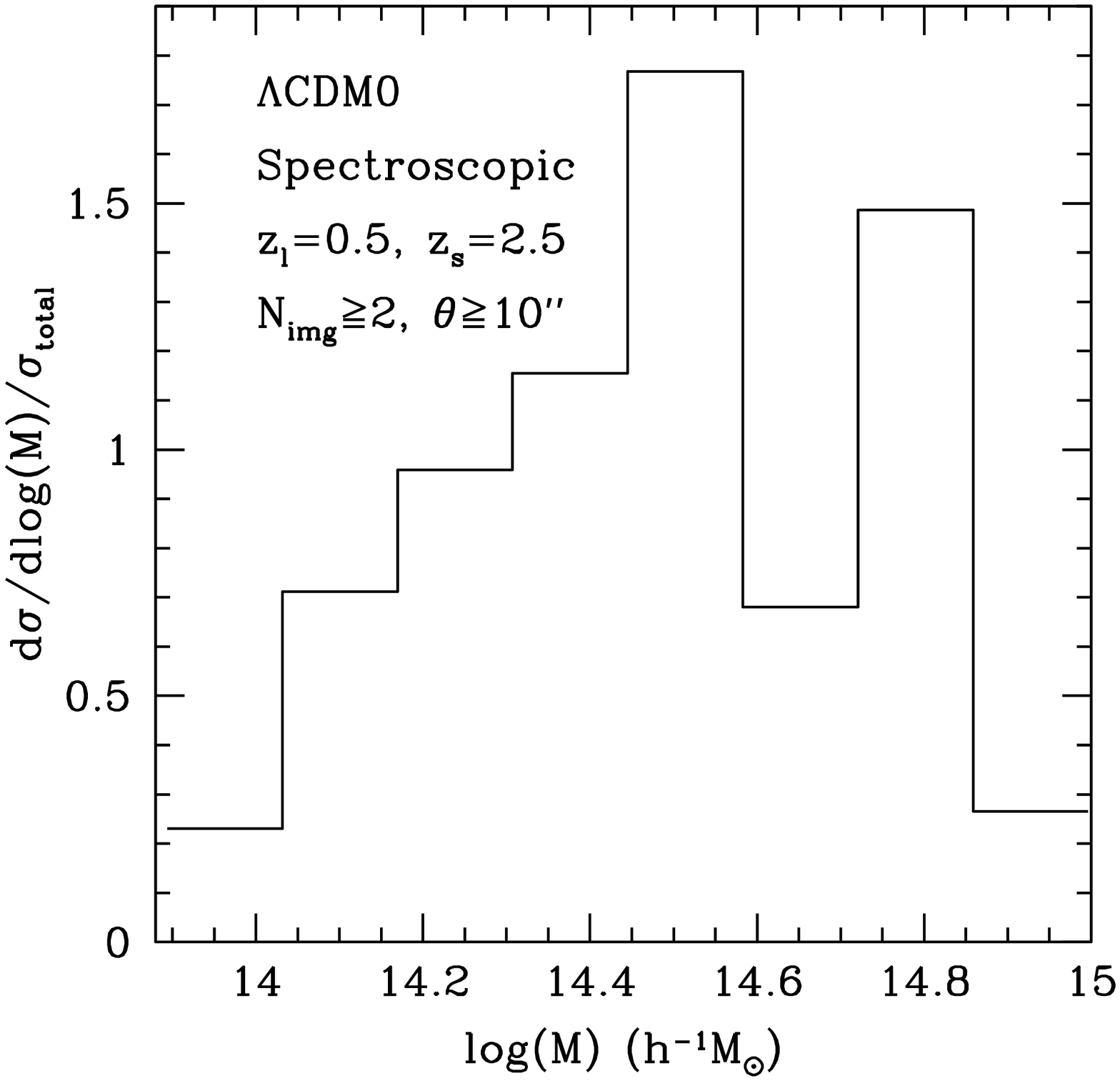}%
\includegraphics[width={\columnwidth}]{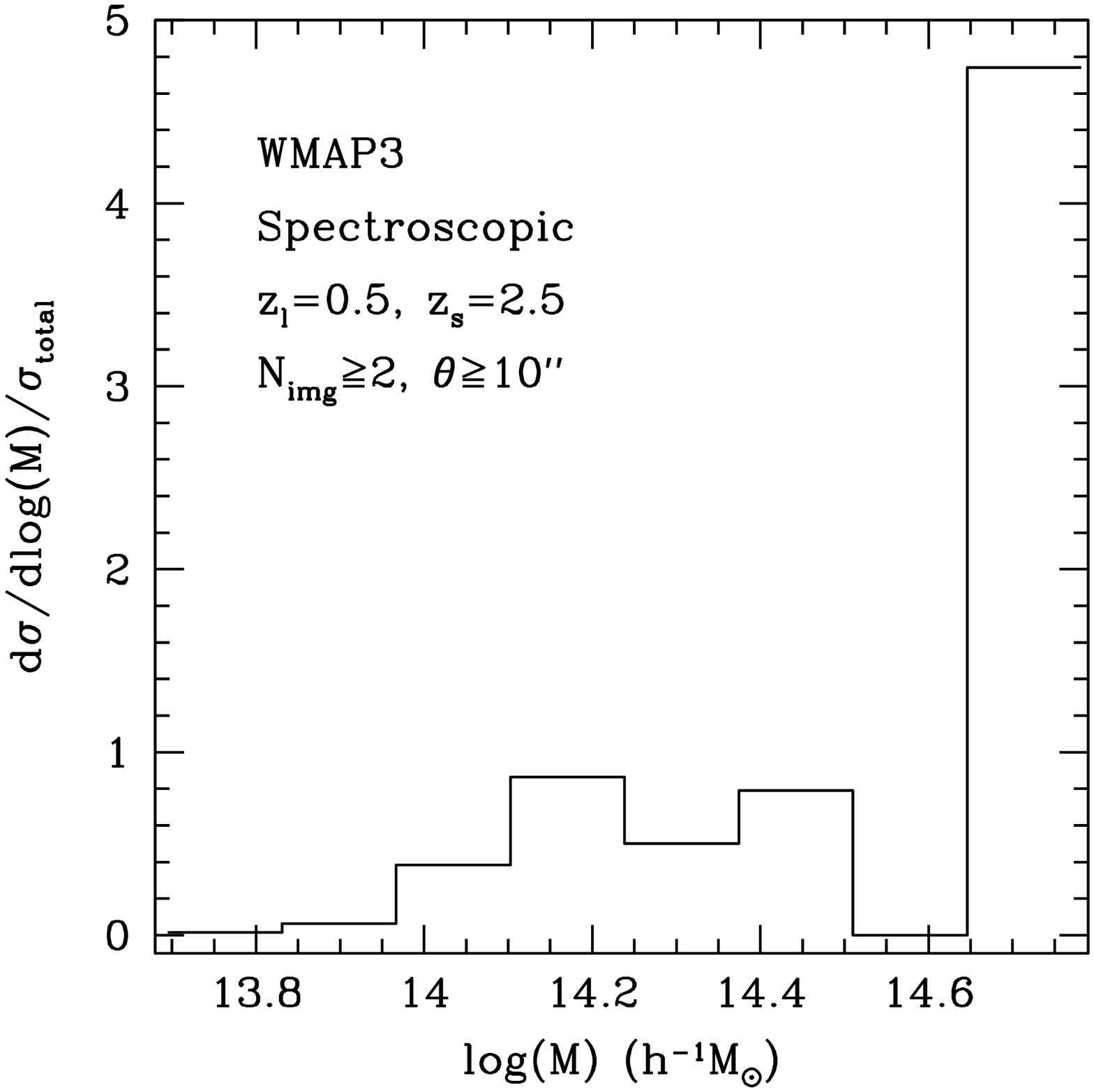}%
}
\caption{Top panel: the solid
and dashed histograms show the mass functions at redshift 0.5 in our $\LCDM 0$
and WMAP3 simulations respectively. The solid and dashed curves are 
the \citet{ST02} mass functions.
The two bottom panels show the differential cross-sections as a 
function of the logarithm of the cluster mass in the two models.
The total area under each curve is normalised to unity. The clusters are at redshift 0.5
and the sources are at redshift 2.5. One can see that the box size of $\LCDM$0 simulation 
is large enough to sample the massive clusters but our WMAP3 simulation does not appear to
have converged in the cross-section as a function of the cluster mass.
} 
\label{fig:variance}
\end{figure}

Our two simulations were run with a box size of $300 h^{-1}\mpc$.
A question naturally arises is whether they sample the mass functions well,
especially for the high-mass end in the WMAP3 model, and how much our results are affected by
the cosmic variance. The top panel in Fig. \ref{fig:variance} shows the mass functions
in our simulations. The solid and dashed histograms are the mass functions at
redshift 0.5 in our $\LCDM$0 and WMAP3 simulations respectively. The solid and dashed curves are
the theoretical \citet{ST02} mass functions, which is based on the Press-Schechter (1974) formalism and the
ellipsoidal collapse model (\citealt{SMT01}). 
We can see that the number density of halos in the WMAP3 model is always 
smaller than that in the $\LCDM$0 model on cluster mass scale.
For $M \sim 10^{14} h^{-1}M_\odot$, the abundance of halos
is lower by a factor of 3.1 in the WMAP3 model compared with that in the
$\LCDM$0 model; for $M \sim 10^{15}h^{-1}M_\odot$, the reduction factor is $\sim 8.8$. 
The number density in the last bin of the WMAP3 mass function is higher
than prediction by a factor of $\sim$ 6 due to the presence of just
6 clusters. The low number of clusters at the high mass end in the
WMAP3 model implies that
ideally a larger simulation box is required to sample at least a 
similar number of massive clusters as in the $\LCDM$0 model.
The two bottom panels show the differential cross-section as a 
function of the cluster mass in the two simulations. The total area 
under each curve is normalised to unity.
The clusters are at redshift 0.5 and the sources are at redshift 2.5.

The mass distribution of SDSS J1004+4112 was studied in \citet{Ota06} 
using X-ray data. Assuming isothermality and an
NFW profile for the cluster, the virial mass was derived to be $
M \sim 4.2^{+2.6}_{-1.5} \times 10^{14} h^{-1}M_\odot$.
The mass of SDSS J1029+2623 is more uncertain. The image separation
implies a cluster velocity dispersion $\sigma \sim \rm 900km/s$ 
(see \citealt{Inada06}), and this can be converted to
 $M \sim 5.6 \times 10^{14} h^{-1}M_\odot$
using cluster scaling relations, but with a large error bar.
The box size of $\LCDM$0 simulation 
appears to be large enough to produce such massive clusters
  easily. On the other hand, there are 
only ~20 clusters in the last three bins in the bottom-right panel of
Fig. \ref{fig:variance}, such a small number may be not sufficiently sample the
clusters well in the mass-concentration space and result in a relatively
large errors of differential cross-section at the very high-mass end. But
the final optical depth is the integration of the  mean cross-section as a
 function of redshift, which will suppress the final variance. 
 However our WMAP3 simulation does
suffer seriously from the cosmic variance
 and appears to have too many very massive
clusters compared to the theoretical prediction. In \citet{Li06b}, we 
ran a lower resolution simulation in
the WMAP3 model which evolves $N_{\rm DM}=512^3$ dark matter particles
in a box with sidelength of $600 h^{-1} \mpc$. Due to the larger volume,
that simulation
better samples the high mass tail of the cluster mass function and merger events.
Comparing these simulations, we found that the smaller simulation over-estimates
the number of giant arcs by a factor of 2. For the same reason, the number of
wide-separation lensed quasars in the WMAP3 may be over-estimated by a similar factor. The WMAP3 cosmology is compatible with observations only at $\sim
4\%$ and $\sim 2\%$ level for an over-estimation by a factor of 1.5 
and 2 respectively. However, we caution that to more properly
account for the cosmic variance, it is necessary to run a much larger
simulation with high resolutions in order to sample the cluster mass
function sufficiently.

\section{Summary and Discussions}

In this paper, we have considered the number of multiply-imaged quasars
lensed by clusters of galaxies in the SDSS photometric and spectroscopic quasar
samples. A similar, earlier attempt was made by \citet{Hen07b}. This study
extends their study by exploring how the predictions depend on the
cosmology. We also study the multiply-imaged quasars as a function of image
multiplicity, and explored in some detail the properties of 
central images. Our main conclusions are as follows
\begin{enumerate}
\item We found that the predicted multiply-imaged quasars with separation
  $>10\arcsec$ in the SDSS photometric sample (with an effective area 8000 deg$^{2}$) is 
about 6.2, and about 2.6 in the spectroscopic
  sample  (with an effective area 5000 deg$^{2}$) in $\LCDM$0 model. 
These numbers are reduced by a factor of $\sim$7 or more
  in the WMAP3 model.
\item The predicted cluster lens peaks around redshift 0.5, and 90\% are
  between 0.2 and 1. This distribution is largely independent of cosmology,
  and similar for both the photometric and spectroscopic samples.
\item The relative number of systems with $\Nimage\ge 4$ images
  and those with $\Nimage \ge2$ images is about 1/3.5. This ratio is largely
  cosmology independent, and similar for both the photometric and
  spectroscopic quasar samples. 
\item Because we modelled the BCGs as a truncated isothermal sphere,
  this creates a region, comparable in size to the angular Einstein radius of the BCG,
  inside which the central image disappears. For most central
  images in five-image configurations are quite faint, and close to the
  cluster centres. For three-image systems, the central images are
  brighter and further away from the cluster centres.
\end{enumerate}

One uncertainty in our calculation is the ad hoc inclusion of a BCG
at the centres of clusters at all redshifts. In reality, the BCGs
are assembled as a function of time, and so it is interesting
to estimate the effects of an evolving population of BCGs. We adopt a simple model
by assuming there are no BCGs at the centres of clusters above redshift
0.5. Hennawi et al. (2007) finds that the optical depth decreases by a factor of
1/3 if all the BCGs are ignored. However, from Fig. 6, the contribution of
clusters with $\zl \ga 0.5$ is about 1/2 of the total cross-section.
Thus if we remove all the BCGs from the clusters above redshift 0.5, the optical
depth will decrease roughly by $\sim 16\%$. So the evolution effect 
may be modest. In any case, such a decrease in the optical depth will
make the match between the WMAP3 model and the observations
slightly worse.

Our predicted numbers of multiply-imaged quasars for both the
photometric and spectroscopic samples are lower than those by \citet{Hen07b}
roughly by a factor of 2 in the $\LCDM$0 cosmology. Part of this difference is due to the slightly
higher $\sigma_8$ (0.95) adopted by \citet{Hen07b}. The 
abundance for typical lensing clusters around $M >10^{14.5}h^{-1} M_\odot$ at redshift
0.5 is about 30\% higher than that in our model with $\sigma_8=0.9$. 
Their higher $\sigma_8$ not only provides more massive clusters but also
makes the cluster formation earlier and with higher concentration; both
may increase the predicted number of cluster lenses. 
Our quasar luminosity function is similar to those
adopted in \citet{Hen07b} for the low redshift intervals, but is a factor of
2 higher than that adopted in their paper for the high-redshift interval
($\zs \sim4$). So the difference in the source luminosity function goes in the
opposite direction. Note, however,
the number of such higher redshift quasars is small and so this should not impact
significantly on the overall predicted large-separation lenses (see
Fig. \ref{fig:source}).

For galaxy-scale lenses, the number of lenses with $\Nimage\ge 4$ is
about one half of that with $\Nimage \ge 2$. For example, the
statistical sample of the Cosmic Lens All Sky Survey (CLASS) has
13 lenses, 6 have at least 4 images (including one with 6 images,
\citealt{Browne03, Chae03}). 
Our study indicates the ratio is about 0.3 for cluster-scale lenses,
 about a factor of 1.5 smaller. This is consistent with
\citet{OK04}. Using a triaxial cluster  
model, they found the ratio is 0.2 for inner density profile 
$\alpha=1$ and 0.4 for $\alpha=1.5$.  
The comparison between the two studies is, however, not straightforward as we use
different definitions of $N_{\rm img}$ and their study also 
all of the faintest images in the multiply imaged systems can be detected, while
in our study we assume certain magnitude limits for the images to be observable 
(see \S\ref{sec:k-corr}). If we relax the magnitude limit to $i=25$, then we find that our
ratio is changed to 0.35, still within the bracket of the two values of
\citet{OK04} for different inner density slopes.
As discussed by
\citet{Rusin01} and \citet{OK04}, this ratio depends on a number of parameters, including
the shape of the lensing potential and the density slopes. 
It appears that this is mainly because of the shallower inner density slope which
causes the factor of 1.5 difference in the relative
numbers between cluster lenses and galaxy lenses.

One striking result of this study is the reduction in the predicted
number of large-separation lenses in the WMAP3 model compared with that
in the $\LCDM$0 model. This reduction is comparable to that for
the giant arcs ($\sim 6$) in these two cosmologies. However, we emphasise that the
uncertainty for the quasar sample is smaller, because the source
population is reasonably well known (see \S\ref{sec:lf}).
There are two lens systems, J0114+4112 and J1029+2623, have been discovered 
in the SDSS spectroscopic sample with an effective area of $\sim 6500$deg$^2$.
In the $\LCDM$0 model, from Fig. \ref{fig:multi}, the number of lenses with
 separation larger than
$14.6\arcsec$ quadrupole-images and $22.5\arcsec$ double images are
0.7 and 1.2. This is roughly consistent with what
 we have discovered in the SDSS. 
However, for the WMAP3 model, the number of lenses with separation larger than
$14.6\arcsec$  quadrupole-images and $22.5\arcsec$ double images are
0.09 and 0.14 for the whole spectroscopic sample. 
 Assuming a Poisson distribution, the probability to observe 
 two quasar systems in the WMAP3 model is $\sim$ 8\% and
 decreases to 4\% or lower if we account for the cosmic variance
 in our simulation (see \S\ref{sec:cosmic_variance}). So it appears that
a higher $\sigma_8$ or a higher $\omega0$ is preferred by the data.

Clearly it is desirable to have a much bigger sample of large-separation
multiply-imaged quasars. Our study shows that most cluster lenses have
already been discovered in the SDSS spectroscopic quasar sample,
but several more candidates may yet to be
discovered in the photometric sample. Future large
surveys may discover many more such examples (\citealt{Wittman06}),
which will provide strong constraints on the cosmology and the central
mass profiles of clusters of galaxies.
\section*{Acknowledgment}
We thank Drs. Neal Dalal and Joseph Hennawi for helpful discussions.
 The referee is thanked for a careful report which improved the paper.
This work is supported by grants from NSFC (No. 10373012, 10533030),
Shanghai Key Projects in Basic research (No. 04JC14079 and
05XD14019). The WMAP3 simulation was performed at the Shanghai Supercomputer Center. SM acknowledges the Chinese Academy of Sciences and the
Chinese National Science Foundation for travel support. This work was also
partly supported by the visitor's grant at Jodrell Bank, 
the Department of Energy contract DE-AC02-76SF00515 and by the
European Community's Sixth Framework Marie 
Curie Research Training Network Programme, Contract No. MRTN-CT-2004-505183 ``ANGLES''.

\end{document}